\documentclass[preprint,12pt,sort&compress]{elsarticle}

\usepackage{amssymb}
\usepackage{amsmath}

\usepackage{booktabs}
\usepackage{multirow}
\usepackage{subcaption}
\usepackage[unicode=true,
bookmarks=true,bookmarksnumbered=true,bookmarksopen=true,bookmarksopenlevel=2,
breaklinks=false,pdfborder={0 0 1},backref=false,colorlinks=true]
{hyperref}

\usepackage{algorithm}
\usepackage{algpseudocodex}

\usepackage{bbm}

\usepackage{array, ragged2e}
\newcolumntype{P}[1]{>{\centering\arraybackslash}p{#1}}

\newcolumntype{C}[1]{>{\centering\arraybackslash\hspace{0pt}}p{#1}}

\newcolumntype{L}[1]{>{\raggedright\arraybackslash\hspace{0pt}}p{#1}}

\journal{Journal of Computational and Applied Mathematics}

\begin{document}
	
\begin{frontmatter}
	
	\title{Thin Plate Spline Surface Reconstruction via the Method of Matched Sections}
	
	\author[1]{Igor Orynyak}
	\ead{igor_orinyak@yahoo.com}
	\author[1]{Kirill Danylenko}
	\ead{k.a.danylenko@gmail.com}
	\author[1]{Danylo Tavrov\corref{cor1}}
	\ead{tavrov.danylo@lll.kpi.ua}
	\cortext[cor1]{Corresponding author}
	
	\affiliation[1]{organization={Applied Mathematics Department, National Technical University of Ukraine “Igor Sikorsky Kyiv Polytechnic Institute”},
		addressline={37 Beresteiskyi Ave}, 
		city={Kyiv},
		postcode={03056},
		country={Ukraine}}
	
	\begin{abstract}
		This paper further develops the Method of Matched Sections (MMS), a robust numerical framework for the solution of boundary value problems governed by partial differential equations. It demonstrates its unique applicability to the challenges of surface modeling, which lie at the intersection of computational mechanics and computer graphics. This work shows how the MMS successfully bridges this gap. By decomposing the domain into an assembly of 1D directional components matched along their entire boundaries, the method inherently enforces the continuity of all variational parameters, including second-order (curvature) and third-order (shear) derivatives. We demonstrate the method's advanced capabilities in high-fidelity surface reconstruction and blending, showing that it consistently generates energetically optimal, fair surfaces even from complex boundary conditions or sparse internal data points. By advancing the application of the MMS, this research establishes it as a powerful, physics-informed geometric tool that satisfies the dual demands of rigorous numerical analysis and aesthetic computer-aided design.
	\end{abstract}
	
	\begin{highlights}
		\item MMS enables robust physics-based surface reconstruction
		\item Method inherently enforces global curvature continuity
		\item Novel regularization handles singular corner point constraints
		\item Bridges gap between computational mechanics and graphics
	\end{highlights}

	
\end{frontmatter}

\section{Introduction}
\label{sec:introduction}

Constrained surface reconstruction is a fundamental problem in computer-aided design (CAD), computer graphics, and reverse engineering. The methodologies for addressing this problem can be broadly divided into two categories: purely geometric and physics-based \cite{Zheng2003}.

The geometrical spline-based surfaces include Bézier, B-splines, and Non-Uniform Rational B-Splines (NURBS) surfaces. These are flexible and can be modified by controlling a set of parameters such as the order of basic functions, control points (number, positions, multiple vertices), and knot vectors (type, distances, multiple knot values, weights) \cite{Rogers2000}. They are very popular, and NURBS surfaces have become a standard in various industries for the digital representation and data exchange of geometric forms. Despite rapid advances in computational techniques, theoretical substantiation, and ready-to-use primitives, this traditional spline-based modeling can be difficult and counter-intuitive \cite{Wang2023}. In such surface reconstruction methods, the resulting shapes often depend on the user's heuristic skills and perceptions of users, leading to inconsistent representations for identical data \cite{Wang2021}. The reason is that geometrical statements are often ill-posed, and additional criteria for the ``fairness'' of the surface are needed. Consequently, surface reconstruction can be considered a constrained optimal approximation problem in two dimensions, which requires a rigorous optimization procedure \cite{Terzopoulos1984}.

Variational techniques are used to optimize the shape of a surface by minimizing a specific energy functional, which often  possesses clear physical analogues \cite{Kovacs2020}. The thin-plate energy functional is mainly used to smooth the curvatures (minimize energy) of Bézier splines \cite{Hu2022} and B-splines \cite{Mosbach2022}. For the task of interpolation, this energy functional, $E$, minimized the aggregate curvature \cite{Terzopoulos1984}:
\begin{equation}
	E(W) = \iint \left( \left(\frac{\partial^2 W}{\partial x^2}\right)^2 + 2\left(\frac{\partial^2 W}{\partial x \partial y}\right)^2 + \left(\frac{\partial^2 W}{\partial y^2}\right)^2 \right)\, dx dy\;,
	\label{eq:energy_functional}
\end{equation}
where $W$ is the scalar field (surface elevation).

Direct physics-based methods have surged in popularity in recent years, especially in computer graphics, where their use has skyrocketed over the past few decades \cite{Holz2025}. This is related to a demand for high-fidelity predictive simulations of real-life behavior \cite{Longva2020}. Their significance is twofold: 1) they satisfy well-posed partial differential formulations from mathematical physics; and 2) these functionals are considered the best for guaranteeing optimal surface fairness. The splines that originate from the elastic thin plate bending theory are very often considered optimal in this regard \cite{Terzopoulos1984, Moreton1992}. Such surface splines are natural 2D extensions of 1D beam splines \cite{Celniker1991}. Similarly, 1D elastic beams provide both the technique of solution and the criteria for fairness \cite{Levien2009, Moreton1992b}.

However, a critical distinction must be made. Like any boundary value problem, the solution must satisfy: a) the governing differential equation, b) internal constraints, and c) boundary conditions. The governing differential equation of the problem is the biharmonic equation:
\begin{equation}
	\Delta^2 W(x,y) = q(x,y)\;,
	\label{eq:biharmonic_equation}
\end{equation}
where $\Delta$ is the Laplace operator, and $q(x,y)$ represents external data terms or constraints. The solution to \eqref{eq:biharmonic_equation} for $q(x,y)=0$ automatically provides the minimum of the energy functional \eqref{eq:energy_functional}.

The term and technique of the thin plate spline (TPS) is very popular \cite{Harder1972, Bookstein2002, Zhao2022}; however, it is often restricted to a linear combination of radial basis functions (RBFs) \cite{Buhmann2000, Carr2001}. These particular solutions,
\begin{equation}
	W_r = A \cdot r^2 \ln r\;,
	\label{eq:1c}
\end{equation}
satisfy \eqref{eq:biharmonic_equation} everywhere except at constraint points, with $r$ being the local radial distance $r$ (counted with respect to the point of force application) and $A$ an unknown constant related to the value of the force.

These functions might have some advantages over their counterparts, but they do not provide a solution to the boundary value problem, which additionally requires the general solution of the homogeneous equation \eqref{eq:biharmonic_equation}. The problem with the application of the radial functions is that they lead to a dense matrix and therefore have a high computational cost. Furthermore, the problem may become ill-conditioned due to the global character of the thin-plate radial functions \cite{Chen2012, Majdisova2017}. The main drawback of the thin-plate radial function method, however, is its inability to account for constraints at the boundaries.

Therefore, it seems natural to resort to the experience gained in the theory of plates. There are two different approaches for the solution of \eqref{eq:biharmonic_equation} that account for constraints and boundaries: analytical and numerical. Generally, analytical solutions lack the universality required for complex boundary conditions. Their applications are restricted to narrower problems, for example, for blending (matching between known surfaces). An approach that provides analytical solutions to harmonic as well as biharmonic problems with respect to some parametric surface coordinates $u$ and $v$ (PDE-based surface design) for simpler geometries was suggested in \cite{Bloor1989}. It was further enhanced for free-form surface generation for different boundary conditions, including periodic solutions \cite{Bloor1990}. This method has found applicability in the presentation of facial blendshapes \cite{Fu2021}, in the optimization design of head shapes for improving the aerodynamic performance of high-speed trains \cite{Wang2021b}, and in the generation of optimal surfaces with curvature-level continuity under the constraint of feature curves for automotive styling design \cite{Wang2023}. Yet the main question regarding the fairness of the generated surface remains open: Is the PDE-based design able to provide the best properties of the surface \eqref{eq:energy_functional}, accounting for the arbitrariness of the non-orthogonal parametric coordinates $u$ and $v$?

To address boundary constraints rigorously, we turn to numerical methods in Computational Mechanics (CM). The finite element  method (FEM) was proposed for surface reconstruction \cite{Terzopoulos1984}. As demonstrated in \cite{Celniker1991}, the usual Ritz-based energy minimization approach with triangular elements is applied to derive the stiffness matrix and forcing vector. The combined use of smoothing techniques and mixed finite element techniques for the thin plate with the application of rectangular elements is given in \cite{Roberts2003}.

Nevertheless, although the essence and governing equations for the thin plate model are the same for both CM and Computer Graphics (CG), they rarely interact in their methods and findings. A rare positive example is in 3D rod deformation, where some findings from the CG community \cite{Bergou2008} are also very popular in the CM world. Furthermore, a comparison between two groups of existing codes has demonstrated that CM codes sometimes (for the spatial rod) perform better than well-known reference models from Mechanical Engineering, yet CG thin plate design codes generally fail to pass some simple tests \cite{Romero2021}. The CG community is still not eager to rely on the findings of CM, perhaps because their primary goals are different. Despite the declared need for accuracy, CG requires, first and foremost, the continuity of solutions, including the second derivatives (curvatures), everywhere within the body. For CM, the main goal is accuracy at specific points.

The main peculiarity of traditional FEM is clearly outlined in a series of works by Almeida et al. \cite{Almeida2020, Almeida2017}. In most, if not all, commercial FEM systems, the conforming formulation is used, which satisfies the constitutive relations and compatibility for the assumed displacement field. However, equilibrium is satisfied only in a weak sense; that is, ``they are not in equilibrium with the body forces and do not have tractions that equilibrate with the static boundary conditions and are not continuous between elements'' \cite{Almeida2017}. The same is noted by the authors of \cite{Olesen2018}, who emphasized that Newton's third law is therefore violated at the boundaries between elements. However, it is precisely the mechanical forces and moments in CM solutions that are proportional to the curvatures. Thus, while being accurate, they do not provide a plausible (continuous) picture of the body deformation. 

In a geometric context, these ``forces'' and ``moments'' correspond directly to surface derivatives and curvatures. Thus, while FEM provides accurate displacements, it may fail to provide a $C^2$-continuous (plausible) representation of the surface.
To bridge the gap between CM and CG communities, one must prove that a numerical solution can accurately recover curvature (moments) and enforce high-order continuity. It is worth demonstrating that a biharmonic function can be restored within the whole area by the FE solution based only on the specified boundary conditions, generated by this biharmonic function. It is also necessary to demonstrate that the application of the boundary conditions generated by an arbitrary function leads to the FE solution, whose value of the functional \eqref{eq:energy_functional} is less than the respective value of \eqref{eq:energy_functional} related to this function. 

The recently proposed Method of Matched Sections (MMS), as a universal multiphysical numerical technique \cite{Orynyak2024, Orynyak2025, Orynyak2025b, Danylenko2024}, possesses both accuracy and continuity for functions and their higher derivatives. Specifically, our solution for thin plates \cite{Orynyak2024, Orynyak2025} is applicable to both mechanical deformation modeling and optimal surface reconstruction. Indeed, demonstrating this dual capability is the primary objective of the present study. Furthermore, while MMS typically constructs continuous positions only along characteristic (skeleton) lines, we propose a specialized ``patch construction'' procedure. This method ensures continuity up to the second-order derivatives, including mixed derivatives.

Due to the novelty of our approach, we restrict ourselves to an initially rectangular domain meshed by rectangular elements. This is not a principal restriction of the method, and the successful treatment of curvilinear boundaries by triangular elements was demonstrated in \cite{Danylenko2024}. 
We show the following capabilities:
\begin{itemize}
	\item A mechanically accurate solution for the thin plate is demonstrated for the case of a point constraint applied at a corner. This is a challenging task because the behavior of the solution near a corner differs significantly from the fundamental solution in the vicinity of an internal point \eqref{eq:1c}; consequently, classical TPS cannot be applied in such instances.
	\item The capacity to recover the same biharmonic function under various combinations of prescribed boundary conditions is a capability that is rarely---if ever, to the authors' knowledge---demonstrated by FEM or other numerical techniques.
	\item When boundary conditions are derived from an arbitrary function, the reconstructed function possesses the optimal energy-minimizing property---a feature typically not demonstrated by other numerical or analytical blending methods.
	\item When only a few constraints (prescribed positions) are derived from an arbitrary function, the surface reconstructed by MMS provides a smooth solution that mimics the function's behavior but is energetically more efficient. To the best of our knowledge, this property has not been previously demonstrated by other surface reconstruction methods.
\end{itemize}

The rest of this paper is structured as follows. Section \ref{sec:mms_theory} provides a brief overview of the Method of Matched Sections as applied to thin plate bending, detailing the analytical relations and the matrix formulation. Section~\ref{sec:examples} presents a series of numerical examples, demonstrating the method's accuracy in handling point constraints, its effectiveness in surface blending for both symmetric and non-symmetric biharmonic functions, and its ability to reconstruct surfaces from sparse internal data points. Finally, Section \ref{sec:conclusion} concludes the paper with a summary of the key findings and their implications.

\section{Thin Plate Bending By MMS}
\label{sec:mms_theory}

The problem statement follows the formulation in \cite{Orynyak2024, Orynyak2025}. MMS reduces Partial Differential Equations (PDEs) to a system of Ordinary Differential Equations (ODEs) via a semi-analytical decomposition. The 2D problem is treated as a coupled system of 1D equations along orthogonal directions. Here, we present the solution in a form compatible with the Transfer Matrix Method (TMM) \cite{Leckie1960, Orynyak2024}.

\subsection{Analytical Relations and Parameters Mapping}

\begin{figure}[ht!]
	\centering
	\begin{subfigure}[b]{0.5\textwidth}
		\centering
		\includegraphics[width=\textwidth]{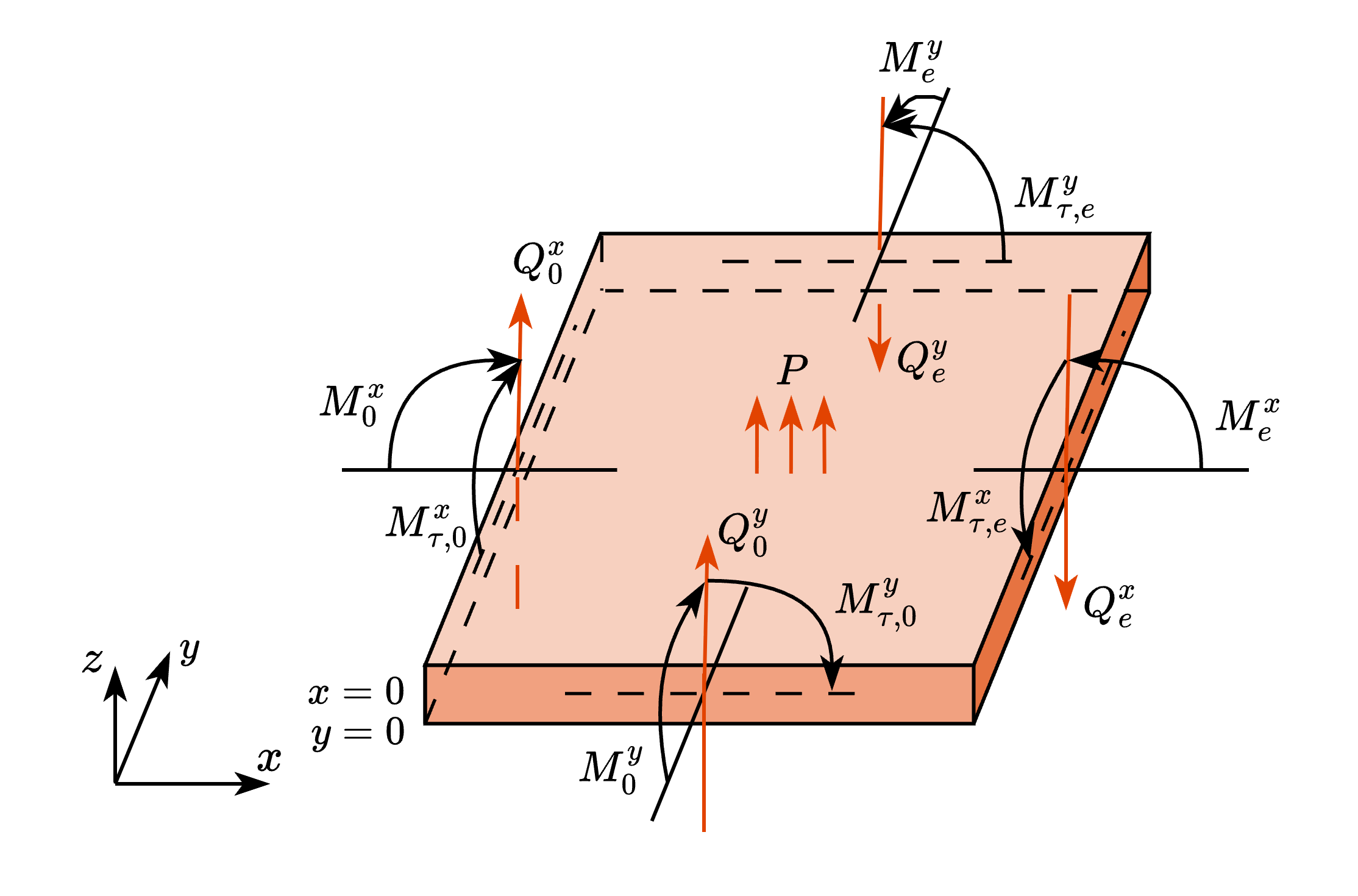}
		\caption{Directions of derivative parameters}
		\label{fig:mms_forces}
	\end{subfigure}
	\hfill
	\begin{subfigure}[b]{0.4\textwidth}
		\centering
		\includegraphics[width=\textwidth]{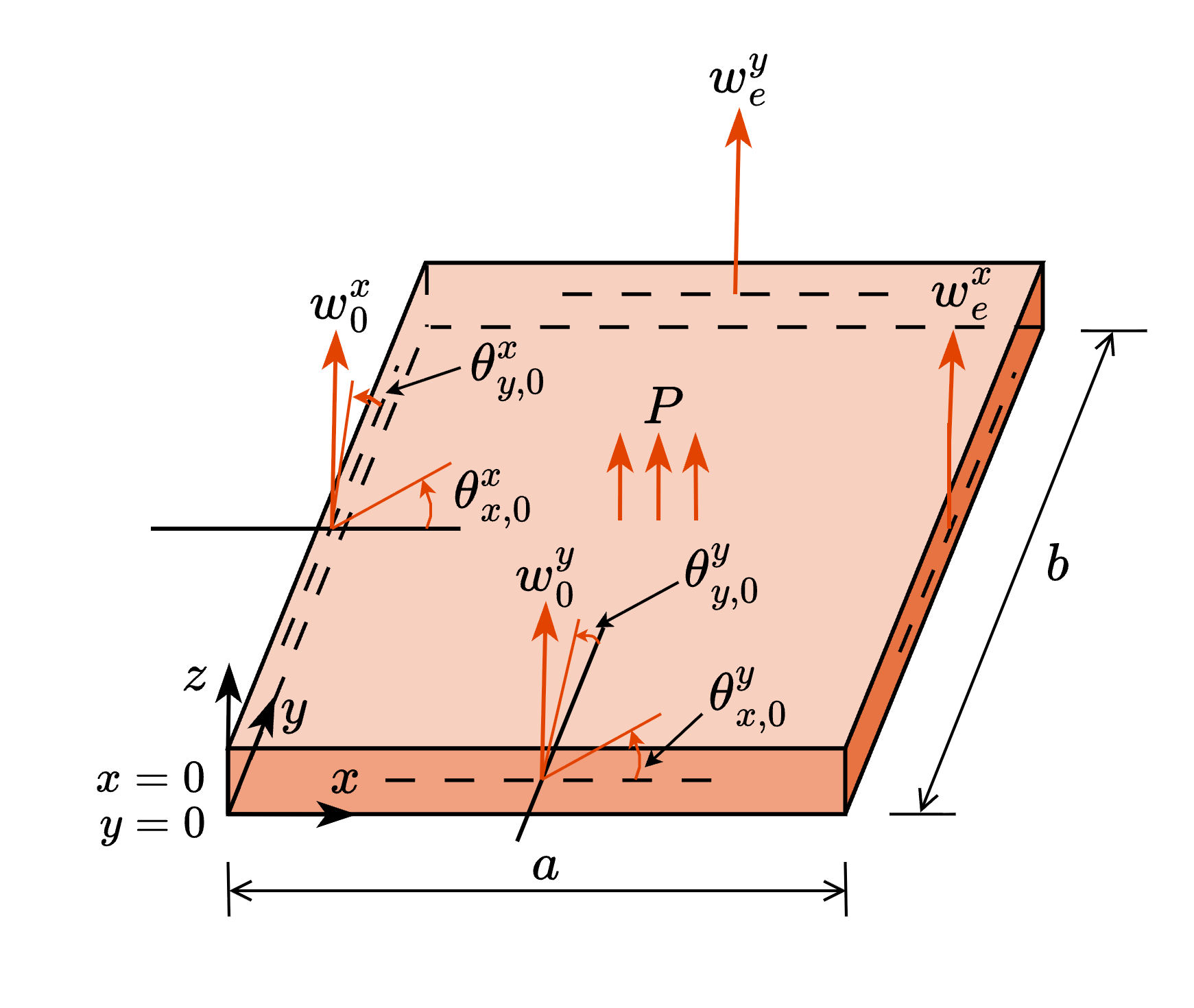}
		\caption{Directions of kinematic parameters}
		\label{fig:mms_kinematics}
	\end{subfigure}
	\caption{General scheme of a finite element decomposition}
	\label{fig:mms_element_scheme}
\end{figure}

Consider the rectangular element shown in Fig.~\ref{fig:mms_element_scheme}, with dimensions $a$ and $b$ along the $x$ and $y$ axes. These parameters govern the accuracy and efficiency of the MMS. The local origin is situated at the lower-left corner. We will give the analytical solution taking into account the Poisson's ratio $\nu$, which characterizes the deformation resulting from a force applied in the perpendicular direction and varies by material in mechanical tasks. For the task of surface reconstruction, it is set to $0$.

The surface element is decomposed into two 1D directional components: the $X$-section (along $y=b/2$) and the $Y$-section (along $x=a/2$). The $X$-section is characterized by 6 parameters, which depend only on $x$ and are related to the central line of this beam $(x, b/2)$. They are:
\begin{itemize}
	\item surface elevation (transverse displacement $w^x(x)$);
	\item normal first order derivative, or normal slope (angle of rotation of the beam $\theta_{n}^x(x)$);
	\item tangent first order derivative, or angle of twisting $\theta_{\tau}^x(x)$;
	\item distributed moments applied to the $X$-section: $M_{n}^x(x)$ (principal curvature parameter, or bending moment) and $M_{\tau}^x(x)$ (twisting curvature parameter, or twisting moment); 
	\item third-order derivative, or distributed transverse force $Q^x(x)$. 
\end{itemize}
Analogously, the $Y$-section is characterized by 6 parameters, which depend only on $y$ and pertain to its central line $(a/2, y)$, namely, $w^y(y)$, $\theta_{n}^y(y)$, $\theta_{\tau}^y(y)$, $M_{n}^y(y)$, $M_{\tau}^y(y)$, and $Q^y(y)$. The superscripts are used to indicate the plane the parameters are related to. 

These 12 parameters are coupled via the following differential and constitutive equations \cite{Orynyak2024}:
\begin{align*}
	\theta_n^x(x) &= \frac{\partial w^x(x)}{\partial x}\;,\quad  \theta_\tau^x(x) = \frac{\partial w^x(x)}{\partial y}\;,\\
	M_n^x(x) &= D \left( \frac{\partial^2 w^x(x)}{\partial x^2} + \nu \frac{\partial^2 w^x(x)}{\partial y^2} \right)\;,\quad  M_\tau^x(x) = D(1-\nu) \frac{\partial^2 w^x(x)}{\partial x \partial y}\;,\\
	Q^x(x) &= D \left( \frac{\partial^3 w^x(x)}{\partial x^3} + \frac{\partial^3 w^x(x)}{\partial x \partial y^2} \right)\;,\\
	\theta_n^y(y) &= \frac{\partial w^y(y)}{\partial y}\;,\quad  \theta_\tau^y(y) = \frac{\partial w^y(y)}{\partial x}\;,\\
	M_n^y(y) &= D \left( \frac{\partial^2 w^y(y)}{\partial y^2} + \nu \frac{\partial^2 w^y(y)}{\partial x^2} \right)\;,\quad  M_\tau^y(y) = D(1-\nu) \frac{\partial^2 w^y(y)}{\partial x \partial y}\;,\\
	Q^y(y) &= D \left( \frac{\partial^3 w^y(y)}{\partial y^3} + \frac{\partial^3 w^y(y)}{\partial y \partial x^2} \right)\;,
\end{align*}
where $D$ is the flexural rigidity of the plate and $\nu$ is Poisson's ratio. In this paper, for computer graphics applications, we set $D = 1$ and $\nu = 0$, unless stated otherwise.

The dependencies of these parameters along the central lines are obtained by sequentially integrating the governing ODEs. This integration introduces external load parameters and auxiliary integration constants (denoted $A_1$, $A_2$, and $A_3$). The resulting closed-form analytical expressions for the kinematic and static variables, culminating in the surface elevation functions $w^x(x)$ and $w^y(y)$, are detailed in \ref{sec:appendix_derivations}.

\subsection{Matrix Equations and Algorithm}

To solve the system, we determine the auxiliary constants $A_1, A_2, A_3$ via compatibility conditions at the element center.
We require the ``welding'' of the two directional components, ensuring continuity of elevation and slope:
\begin{align}
	w^x\left(\frac{a}{2}\right) &= w^y\left(\frac{b}{2}\right)\;, \label{eq:coupling_w} \\
	\theta_{n}^x\left(\frac{a}{2}\right) &= \theta_{\tau }^y\left(\frac{b}{2}\right)\;, \label{eq:coupling_theta_n} \\
	\theta_{\tau }^x\left(\frac{a}{2}\right) &= \theta_{n}^y\left(\frac{b}{2}\right)\;. \label{eq:coupling_theta_tau}
\end{align}
These equations effectively substitute the PDE with a coupled ODE system.
The relationship between constants and the state vector $\mathbf{Z}_0$ (the 12 boundary parameters) is expressed as:
\begin{equation}
	\begin{pmatrix} A_1 \\ A_2 \\ A_3 \end{pmatrix} = 
	\begin{pmatrix}
		\alpha_{1,1}&\alpha_{1,2}&\ldots&\alpha_{1,11}&\alpha_{1,12}\\\alpha_{2,1}&\alpha_{2,2}&\ldots&\alpha_{2,11}&\alpha_{2,12}\\\alpha_{3,1}&\alpha_{3,2}&\ldots&\alpha_{3,11}&\alpha_{3,12}
	\end{pmatrix}
	\begin{pmatrix} Z_{1,0} \\ \vdots \\ Z_{12,0} \end{pmatrix} + P \begin{pmatrix} \beta_1 \\ \beta_2 \\ \beta_3 \end{pmatrix}\;,
	\label{eq:coupling_matrix}
\end{equation}
where all coefficients $\alpha_{i,j}$ and $\beta_i$ are known. For convenience, we renumber all 12 initial conditions as a state vector $\mathbf{Z}_0$ with the following coordinates:
\begin{align}
	w_{0}^x      &{}= Z_{1,0}\;,  &\  \theta_{n,0}^x    &{}= Z_{2,0}\;,  &\  \theta_{\tau,0}^x &{}= Z_{3,0}\;, \\
	M_{n,0}^x    &{}= Z_{4,0}\;,  &\  M_{\tau,0}^x     &{}= Z_{5,0}\;,  &\  Q_{0}^x            &{}= Z_{6,0}\;, \label{eq:state_vector_x} \\
	w_{0}^y      &{}= Z_{7,0}\;,  &\  \theta_{n,0}^y    &{}= Z_{8,0}\;,  &\  \theta_{\tau,0}^y &{}= Z_{9,0}\;, \\
	M_{n,0}^y    &{}= Z_{10,0}\;, &\  M_{\tau,0}^y     &{}= Z_{11,0}\;, &\  Q_{0}^y            &{}= Z_{12,0}\;. \label{eq:state_vector_y}
\end{align}
Then it is possible to separately compile the equations for all six parameters that characterize the $X$-section for any $x=\text{const}$. They can be formally presented as:
\begin{equation}
	\begin{pmatrix} Z_1(x) \\ \vdots \\ Z_6(x) \end{pmatrix} = 
		\begin{pmatrix}
		a_{1,1}(x)&\ldots&a_{1,12}(x)\\\vdots&\ddots&\vdots\\a_{6,1}(x)&\ldots&a_{6,12}(x)
	\end{pmatrix}
	\begin{pmatrix} Z_{1,0} \\ \vdots \\ Z_{12,0} \end{pmatrix} + P \begin{pmatrix} b_{P,1}(x) \\ \vdots \\ b_{P,6}(x) \end{pmatrix}\;,
	\label{eq:ftm_x}
\end{equation}
where $Z_1(x) = w^x(x)$ and so on. The coefficients are such that $a_{m,m}(0) = 1$, and all other coefficients are equal to zero at point $x=0$, i.e. $a_{m,k}(0) = 0$, and all $b_{P,m}(0) = 0$. 

For the six parameters that characterize the state of the $Y$-section, it can be written by analogy:
\begin{equation}
	\begin{pmatrix} Z_7(y) \\ \vdots \\ Z_{12}(y) \end{pmatrix} = 
	\begin{pmatrix}
		c_{1,1}(y)&\ldots&c_{1,12}(y)\\\vdots&\ddots&\vdots\\c_{6,1}(y)&\ldots&c_{6,12}(y)
	\end{pmatrix}
	\begin{pmatrix}  Z_{1,0} \\ \vdots \\ Z_{12,0} \end{pmatrix} + P \begin{pmatrix} d_{P,1}(y) \\ \vdots \\ d_{P,6}(y) \end{pmatrix}\;,
	\label{eq:ftm_y}
\end{equation}
where $Z_7(y) = w^y(y)$ and so on. The coefficients are such that $c_{m,m+6}(0) = 1$, and all other coefficients are equal to zero at point $y=0$, i.e. $c_{m,k}(0) = 0$, and all $d_{P,m}(0) = 0$. 

According to the logic of TMM, it is convenient to formally specify a set of 12 auxiliary unknown constants (to be eliminated at the further steps of the calculation process), which are the values of the main parameters of the two 1D components at their ends. For the $X$-section ($x=a$):
\begin{equation}
	\begin{pmatrix} Z_{13} \\ \vdots \\ Z_{18} \end{pmatrix} = 
	\begin{pmatrix}
		a_{1,1}(a)&\ldots&a_{1,12}(a)\\\vdots&\ddots&\vdots\\a_{6,1}(a)&\ldots&a_{6,12}(a)
	\end{pmatrix}
	\begin{pmatrix}  Z_{1,0} \\ \vdots \\ Z_{12,0} \end{pmatrix} + P \begin{pmatrix} b_{P,1}(a) \\ \vdots \\ b_{P,6}(a) \end{pmatrix}\;,
	\label{eq:outlet_x}
\end{equation}
where $Z_{13} = w^x\left(x=a\right)\equiv w_e^x$ and so on, with the subscript $e$ meaning the value of the specific parameter at the end of the section.

For the $Y$-section ($y=b$), we have:
\begin{equation}
	\begin{pmatrix} Z_{19} \\ \vdots \\ Z_{24} \end{pmatrix} = 
	\begin{pmatrix}
		c_{1,1}(b)&\ldots&c_{1,12}(b)\\\vdots&\ddots&\vdots\\c_{6,1}(b)&\ldots&c_{6,12}(b)
	\end{pmatrix}
	\begin{pmatrix}  Z_{1,0} \\ \vdots \\ Z_{12,0} \end{pmatrix} + P \begin{pmatrix} d_{P,1}(b) \\ \vdots \\ d_{P,6}(b) \end{pmatrix}\;,
	\label{eq:outlet_y}
\end{equation}
where $Z_{19} = w^y\left(y=b\right)\equiv w_e^y$ and so on.

Equations \eqref{eq:ftm_x} and \eqref{eq:ftm_y} are usually called field transfer matrix (FTM) equations \cite{Leckie1960}, which map the initial (inlet) state to any point $(x, y)$. For the upper and right boundaries (outlets), equations \eqref{eq:outlet_x} and \eqref{eq:outlet_y} relate the outlet states to the inlet parameters.

The above connection equations \eqref{eq:outlet_x} and \eqref{eq:outlet_y} relate to a single element. In this case, we have 24 unknowns ($Z_1$--$Z_{24}$), which are related by 12 connection equations. They must be supplemented by 3 boundary conditions at each of the 4 sides. Thus, the total number of unknowns is 24, which is equal to the number of equations.

For a multi-element mesh, we need to provide continuity at the borders between elements. If elements $n_1$ and $n_2$ share a boundary, designated as $j_1$ and $j_2$ respectively, then their state vectors must match:
\begin{equation}
	\mathbf Z_{i,j_1}^{n_1} = \mathbf Z_{i,j_2}^{n_2}\;, \quad 1 \le i \le 6\;.
	\label{eq:conjugation}
\end{equation}
This provides 6 conjugation equations (in TMM terms, a point transfer matrix \cite{Leckie1960}). Each side of an element can thus contain either 3 boundary equations or 3 conjugation equations; overall, the number of boundary or conjugation equations is 12. Thus, each element has 24 unknown parameters for which $12+12$ equations exist, so the problem can be solved. This enforces $C^2$ continuity or higher (depending on the solved variables), which is a distinct advantage over standard $C^0$ or $C^1$ FEM approaches.

\subsection{Computing Surface Segments Between the Skeleton Lines}
\label{subsec:computing}

Up to this point, the MMS formulation has solved for the structural state vector strictly along the orthogonal skeleton lines of the mesh. While this provides a highly accurate, $C^3$-continuous framework of 1D directional components, the primarily calculated displacements do not constitute a globally continuous 2D function across the entire domain.

Therefore, completing the surface requires interpolating the interior domains between these lines. Because the MMS determines not only the continuous displacements along the skeleton lines but also their $C^2$-continuous transverse rotations, we can accurately approximate the surface in the immediate vicinity of these lines. For small deviations $y = \epsilon_y$ and $x = \epsilon_x$ (where $\epsilon_y, \epsilon_x$ are small parameters), the surface equation is approximated as follows:
\begin{equation}
	\begin{split}
		W^x\left(x,y\right) &= w^x\left(x\right) + \theta_\tau^x\left(x\right)y\;,\\
		W^y\left(x,y\right) &= w^y\left(y\right) + \theta_\tau^y\left(y\right)x\;.		
	\end{split}
	\label{eq:12}
\end{equation}

It should be emphasized that all functions in \eqref{eq:12} maintain at least $C^2$ continuity. Moreover, at the intersections of the vertical and horizontal skeleton lines, both surfaces in \eqref{eq:12} yield identical displacements and rotations. The availability of analytical solutions for these values along the skeleton lines facilitates the construction of continuous surfaces within each rectangular patch.

Therefore, to complete the surface between the skeleton lines, the standard bilinear Coons patches \cite[p. 93]{Salomon} are used, which were previously employed in \cite{Orynyak2025} to model natural vibration modes. The Coons patch interpolates a surface bounded by four curves, ensuring $C^0$ continuity across all boundaries. Note that the $C^2$ continuity of all four boundary functions enables the development of smoother and more efficient patches. Furthermore, the merging and division procedure \cite{Orynyak2024} allows for the decomposition of elements into multiple sub-elements, thereby generating additional skeleton lines. In this study, Coons patches are utilized specifically for cases involving coarse meshes with very few elements.

\begin{figure}[ht!]
	\centering
	\includegraphics[width=0.5\textwidth]{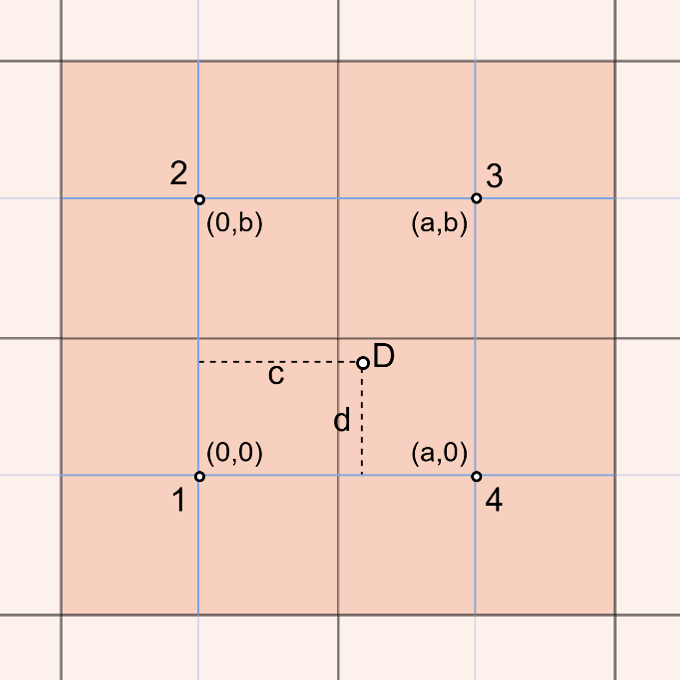}
	\caption{Coons patch for an inner point between four skeleton lines}
	\label{fig:coons}
\end{figure}

Consider the patch bounded by two vertical and two horizontal skeleton lines, as shown in Fig.~\ref{fig:coons}. We designate the intersection points as 1, 2, 3, and 4. Due to the solution procedure \eqref{eq:displacement_x}--\eqref{eq:displacement_y} and the inter-element continuity conditions \eqref{eq:conjugation}, displacements along these four lines maintain $C^3$ continuity, including at the vertices. Denoting the displacements along these boundaries by $w_{1-2}^y\left(y\right)$, $w_{3-4}^y\left(y\right)$, $w_{1-3}^x\left(x\right)$, $w_{2-4}^x\left(x\right)$, and the displacements in points 1--4 by $w_1$--$w_4$, the displacement at an arbitrary point $D(c,d)$ is given by the bilinear Coons formula:
\begin{equation}
\begin{split}
	W^{\text{MMS}}(c, d) &= \frac{b-d}{b} w_{1-3}^x(c) + \frac{d}{b} w_{2-4}^x(c) + \frac{a-c}{a} w_{1-2}^y(d) + \frac{c}{a} w_{3-4}^y(d)\\
	&- \frac{(a-c)(b-d)}{ab} w_1 - \frac{(a-c) d}{ab} w_2 - \frac{c(b-d)}{ab} w_3 - \frac{cd}{ab} w_4\;.
\end{split}
\label{eq:Coons}
\end{equation}
This bilinear Coons patch exactly reproduces the prescribed boundary curves, with the bilinear correction term eliminating corner redundancies. This formulation is directly applicable to interior patches bounded on all sides by skeleton lines.

\begin{figure}[ht!]
	\centering
	\includegraphics[width=0.5\textwidth]{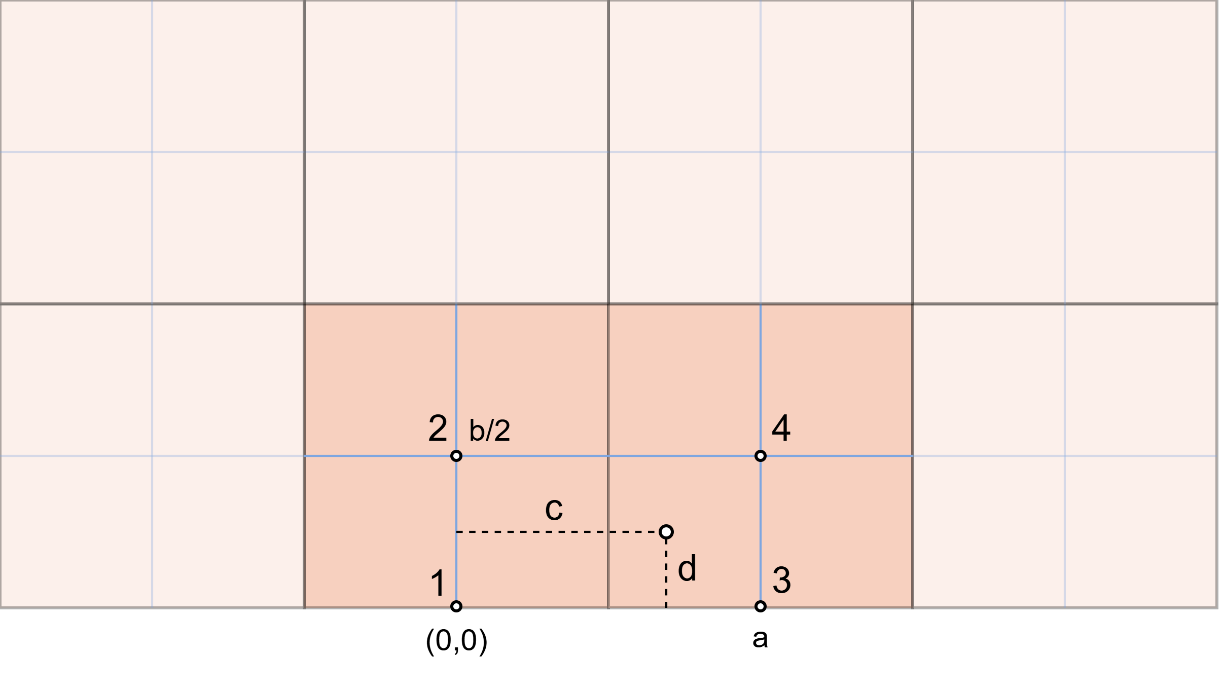}
	\caption{Coons patch for a line lying on the lower outer boundary}
	\label{fig:coonsportion}
\end{figure}

For patches adjacent to an outer boundary, where one bounding curve is not a skeleton line, the missing boundary values are reconstructed using the displacement and tangential rotation from the nearest parallel skeleton lines via \eqref{eq:12}. Consider an element containing a portion of the domain's lower boundary (Fig.~\ref{fig:coonsportion}). While the curve 1--3 (located at $y=0$) is not a skeleton line, both displacements and transverse rotations are available at that boundary from the perpendicular skeleton lines 1--2 and 3--4. By interpolating between these lines via \eqref{eq:12}, the missing boundary curve is reconstructed as:
\begin{equation}
	\begin{split}
		w_{1-3}^x\left(c\right) &= \left(1-\frac{c}{a}\right)\left(w_{1-2}^y\left(0\right) + \theta_{\tau,3-4}^y\left(0\right)c\right) \\
		&+ \frac{c}{a}\left(w_{3-4}^y\left(a\right) + \theta_{\tau,3-4}^y\left(a\right)\left(a-c\right)\right)\;.
	\end{split}
	\label{eq:Coonsouter}
\end{equation}
This equation provides a continuous approximation for the boundary curve 1--3, which is subsequently substituted into the Coons patch formula \eqref{eq:Coons} to determine the displacement at any point within the boundary patch. This approach generalizes to all domain boundaries; in each case, the missing curves are reconstructed from the displacements and transverse rotations of the nearest parallel skeleton lines using \eqref{eq:12}.

In corner elements, displacements at the boundary edge centers are provided by the skeleton lines. From these points, we extrapolate toward the corner using \eqref{eq:12} to obtain two independent displacement estimates, which are then averaged to define the corner value. With the corner displacement established, the two missing boundary curves are reconstructed following \eqref{eq:Coonsouter}, allowing the Coons patch \eqref{eq:Coons} to interpolate the displacement field across the element interior.

\subsection{Regularization of Singular Point Constraints}
\label{subsec:reg}

In surface reconstruction, particularly for interpolation tasks, it is often necessary to constrain specific internal points to exact target elevations. In the biharmonic formulation, a concentrated point constraint acts mathematically analogous to a concentrated transverse force applied to a thin plate.

According to classical plate theory, a concentrated force introduces a curvature singularity. Within the MMS framework, constraining the displacement $w$ at a single element's center induces a sharp gradient in the third derivative (analogous to shear force) between the constrained element and its unconstrained neighbors. While this behavior is mechanically accurate for a true point load, it can produce visually undesirable ``pinching'' or third-derivative discontinuities. This detracts from the global $C^2$ fairness typically demanded in computer graphics applications.

To mitigate this effect and produce visually fair surfaces, we introduce a localized regularization technique. Rather than enforcing the variational constraint strictly at a single element, we distribute it over a local neighborhood of adjacent elements using a smoothing kernel, $\alpha$:
\begin{equation}
	\alpha = \frac{1}{1 + \left( \frac{d}{\zeta \cdot \min(l_x, l_y)} \right)^2}\;,
	\label{eq:force_spreading}
\end{equation}
where $d$ is the Euclidean distance from the center of the adjacent element to the exact attachment point, $l_x$ and $l_y$ are the dimensions of the local elements, and $\zeta$ is a user-defined dimensionless regularization parameter that controls the broadness of the distribution.

To preserve the computational efficiency and local character of the method, the smoothing kernel $\alpha$ is strictly truncated. It is set to zero if the distance between the element centers exceeds a specified threshold, defined here as $d > \frac{\min(l_x, l_y)}{5}$.

By distributing the constraint, the MMS effectively replaces a singular point load with a localized, smooth pressure distribution. Adjusting the regularization parameter $\zeta$ allows for precise control over this smoothing effect, enabling the seamless reconstruction of energetically optimal surfaces that interpolate sparse data points without generating curvature discontinuities.

\section{Numerical Validation and Surface Reconstruction}
\label{sec:examples}

To evaluate the method's geometric efficiency, we calculate the discrete energy integral
\begin{equation}
	E \approx \sum_{n=1}^{N} E_n\;,
	\label{eq:discrete_energy_sum}
\end{equation}
where
\begin{equation}
	\begin{split}
		E_n &= \left( \left(M_{n}^x\left(\frac{a_n}{2}\right)\right)^2 + \left(M_{\tau}^x\left(\frac{a_n}{2}\right)\right)^2 + \left(M_{\tau}^y\left(\frac{b_n}{2}\right)\right)^2 + \left(M_{n}^y\left(\frac{b_n}{2}\right)\right)^2 \right) \\
		&\times (a_n \cdot b_n)\;,
	\end{split}
	\label{eq:discrete_energy_element}
\end{equation}
where $E_n$ is the element energy contribution based on the squared curvature parameters.

\begin{figure}[ht!]
	\centering
	\begin{subfigure}[b]{0.46\textwidth}
		\centering
		\includegraphics[width=\textwidth]{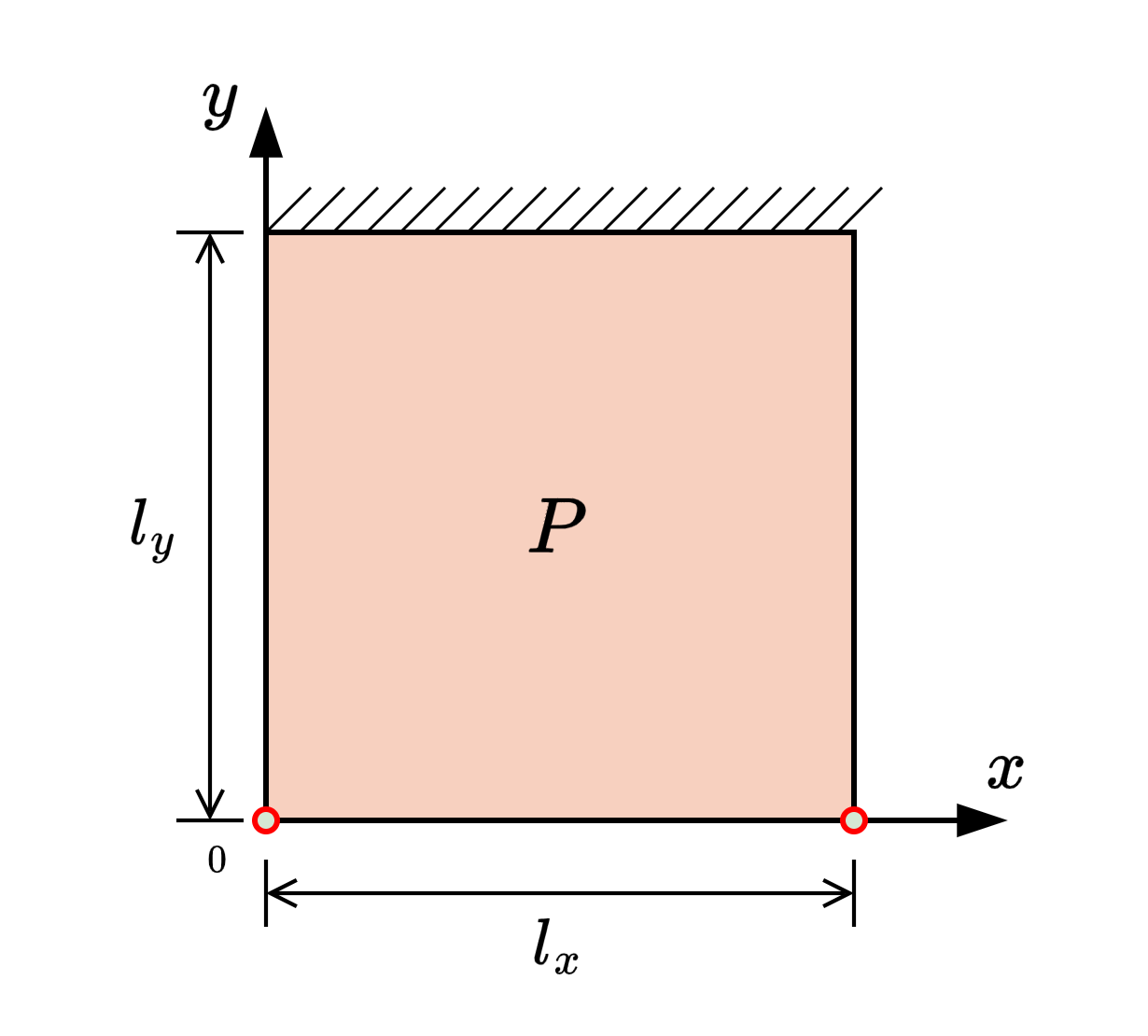}
		\caption{Geometry}
		\label{fig:cfff_geometry}
	\end{subfigure}
	\hfill
	\begin{subfigure}[b]{0.5\textwidth}
		\centering
		\includegraphics[width=\textwidth]{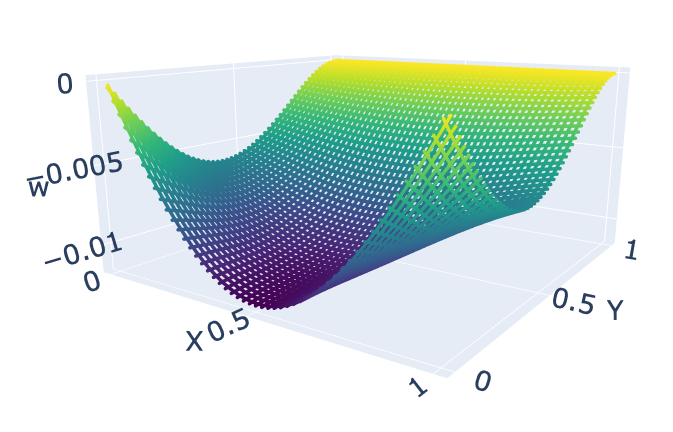}
		\caption{Surface deformation}
		\label{fig:cfff_deformation}
	\end{subfigure}
	\caption{Square domain with singular corner constraints}
	\label{fig:cfff_plate}
\end{figure}

\subsection{Constraint Handling: The Corner-Supported Surface}

a square domain of dimensions $l_x \times l_y$ subjected to a uniform distributed load $q$. The plate has two adjacent corners supported (fixed to zero elevation)  (Fig.~\ref{fig:cfff_plate}). This tests the method's ability to handle singular geometric constraints under uniform forcing.

\begin{figure}[ht!]
	\centering
	\includegraphics[width=0.4\textwidth]{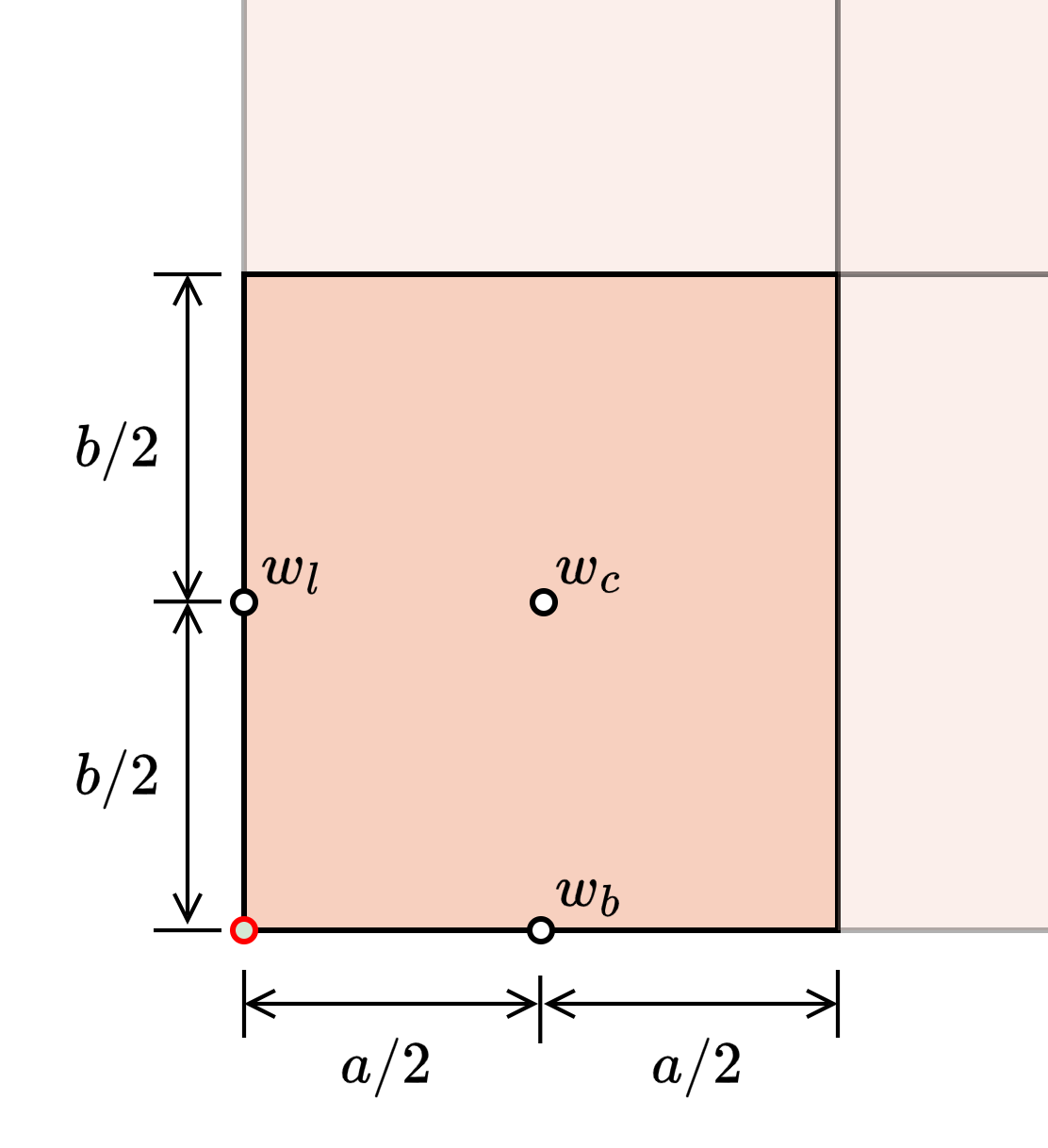}
	\caption{Corner element topology}
	\label{fig:corner_element}
\end{figure}

The MMS enables flexible implementation of corner constraints. We compare three formulations (Fig.~\ref{fig:corner_element}):
\begin{itemize}
	\item \textit{MMS-B}: zero elevation at boundary centers;
	\item \textit{MMS-BA}: incorporating tangential slope constraints;
	\item \textit{MMS-BAM}: incorporating slope and curvature/moment constraints.
\end{itemize}

For the thin plate problem, MMS requires 3 boundary conditions per side. The first approach (MMS-B) involves setting fixed elevations at the center of the boundary side. For the left and lower sides, this gives:
\begin{gather}
	w_{0}^x = Z_1 = 0\;, \quad M_{n,0}^x = Z_4 = 0\;, \quad M_{\tau,0}^x = Z_5 = 0\;, \label{eq:mmsb_left} \\
	w_{0}^y = Z_7 = 0\;, \quad M_{n,0}^y = Z_{10} = 0\;, \quad M_{\tau,0}^y = Z_{11} = 0\;. \label{eq:mmsb_lower}
\end{gather}
The second approach (MMS-BA) considers the tangential angles at the boundary fixation. Taking into account that the position in the corner point is given, we can write for the left-lower point:
\begin{equation}
	w_{0}^x - \theta_{\tau,0}^x \cdot \frac{b}{2} = 0\;, \quad w_{0}^y - \theta_{\tau,0}^y \cdot \frac{a}{2} = 0\;.
	\label{eq:mmsba_corner}
\end{equation}
This leads to the following boundary conditions:
\begin{gather}
	Z_1 - Z_3 \frac{b}{2} = 0\;, \quad Z_4 = 0\;, \quad Z_5 = 0\;, \label{eq:mmsba_left} \\
	Z_7 - Z_9 \frac{a}{2} = 0\;, \quad Z_{10} = 0\;, \quad Z_{11} = 0\;. \label{eq:mmsba_lower}
\end{gather}
In the third approach (MMS-BAM), we also simulate the variational effect of a concentrated Lagrange multiplier at the corner. Assuming we have distributed tangential second derivative and third derivative, we calculate the resulting moment at the corner point and equate it to zero:
\begin{equation}
	M_{\tau,0}^x b - Q_{0}^x b \frac{b}{2} = 0\;, \quad M_{\tau,0}^y a - Q_{0}^y a \frac{a}{2} = 0\;.
	\label{eq:mmsbam_corner}
\end{equation}
This yields the following equations for each side:
\begin{gather}
	Z_1 - Z_3 \frac{b}{2} = 0\;, \quad Z_4 = 0\;, \quad Z_5 - Z_6 \frac{b}{2} = 0\;, \label{eq:mmsbam_left} \\
	Z_7 - Z_9 \frac{a}{2} = 0\;, \quad Z_{10} = 0\;, \quad Z_{11} - Z_{12} \frac{a}{2} = 0\;. \label{eq:mmsbam_lower}
\end{gather}
Table \ref{tab:cfff_comparison_approaches} compares the results against high-precision reference solutions from \cite{Li2015}. Evidently, the third approach gives the best results, but they are still far from the accurate ones.

\begin{table}[ht!]
	\centering
	\caption{Comparison for square domain with singular corner constraints for different MMS approaches ($\nu=0.3$, mesh: $171 \times 171$)}
	\label{tab:cfff_comparison_approaches}
	\resizebox{\textwidth}{!}{%
		\begin{tabular}{lcccccc}
			\toprule
			Approach & $\frac{DW\left(\frac{l_x}{2}, \frac{l_y}{2}\right)}{ql_x^4}$ & $\frac{M_x\left(\frac{l_x}{2}, \frac{l_y}{2}\right)}{ql_x^2}$ & $\frac{M_y\left(\frac{l_x}{2}, \frac{l_y}{2}\right)}{ql_x^2}$ & $\frac{DW\left(\frac{l_x}{4}, \frac{l_y}{2}\right)}{ql_x^4}$ & $\frac{M_x\left(\frac{l_x}{4}, \frac{l_y}{2}\right)}{ql_x^2}$ & $\frac{M_y\left(\frac{l_x}{4}, \frac{l_y}{2}\right)}{ql_x^2}$ \\
			\midrule
			MMS-B & 0.007936 & 0.035962 & 0.050924 & 0.007280 & 0.027425 & 0.054210 \\
			MMS-BA & 0.008026 & 0.037794 & 0.049389 & 0.007293 & 0.028897 & 0.052624 \\
			MMS-BAM & 0.008064 & 0.037878 & 0.049415 & 0.007328 & 0.028983 & 0.052667 \\
			Exact \cite{Li2015} & 0.008052 & 0.037900 & 0.049495 & 0.007316 & 0.029028 & 0.052692 \\
			\bottomrule
		\end{tabular}
	}
\end{table}

\begin{figure}[ht!]
	\centering
	\includegraphics[width=0.7\textwidth]{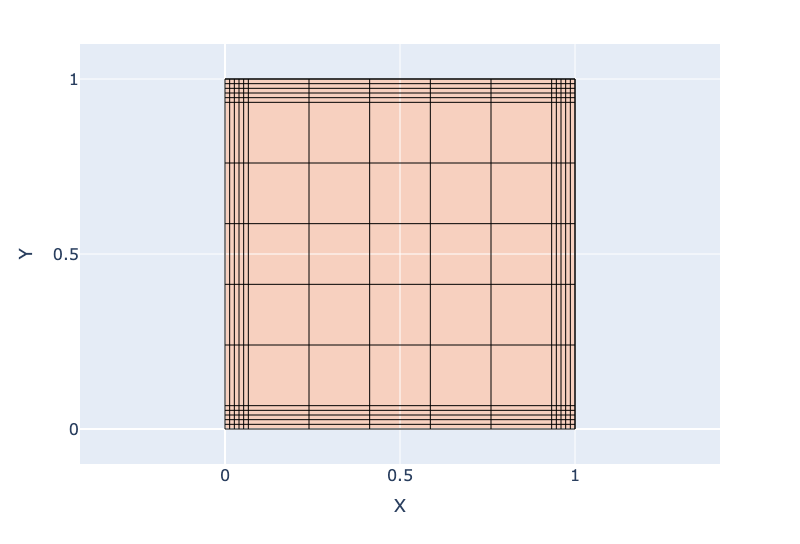}
	\caption{$15 \times 15$ mesh with boundary scaling}
	\label{fig:boundary_scaling_mesh}
\end{figure}

To further refine the result near the singularity, we can either use a much denser mesh or apply local meshing, for example, by resorting to our merging and division procedure \cite{Orynyak2024}. We use a simplified approach instead by applying a linear reduction to several boundary rows of elements, as shown in Fig.~\ref{fig:boundary_scaling_mesh} (where the first and last 4 rows are reduced so that the symmetry of the grid is preserved and all the key points fall on the MMS skeleton lines). This scaling allows us to specify corner boundary conditions much more accurately, improving the overall accuracy without increasing global computation cost (Table \ref{tab:cfff_comparison_scaling}).

\begin{table}[ht!]
	\centering
	\caption{Comparison for square domain with singular corner constraints  for MMS-BAM approach using boundary scaling ($\nu=0.3$, mesh: $251 \times 251$)}
	\label{tab:cfff_comparison_scaling}
	\resizebox{\textwidth}{!}{%
	\begin{tabular}{cc cc cc cc}
		\toprule
		\multirow{2}{*}{$y$} & \multirow{2}{*}{$x$} & \multicolumn{2}{c}{$DW/ql_x^4$} & \multicolumn{2}{c}{$M_x/ql_x^2$} & \multicolumn{2}{c}{$M_y/ql_x^2$} \\
		\cmidrule(lr){3-4} \cmidrule(lr){5-6} \cmidrule(lr){7-8}
		& & Exact \cite{Li2015} & MMS & Exact \cite{Li2015} & MMS & Exact \cite{Li2015} & MMS \\
		\midrule
		0 & $0.25l_x$ & 0.008098 & 0.008102 & 0.076216 & *0.076467 & 0 & 0 \\
		& $0.5l_x$ & 0.011237 & 0.011241 & 0.094329 & *0.094558 & 0 & 0 \\
		\addlinespace
		$0.25l_y$ & 0 & 0.005163 & 0.005165 & 0 & 0 & 0.071900 & *0.072110 \\
		& $0.25l_x$ & 0.008934 & 0.008937 & 0.050837 & 0.050839 & 0.056157 & 0.056152 \\
		& $0.5l_x$ & 0.010596 & 0.010598 & 0.066909 & 0.066911 & 0.052390 & 0.052379 \\
		\addlinespace
		$0.5l_y$ & 0 & 0.005797 & 0.005798 & 0 & 0 & 0.062840 & *0.062970 \\
		& $0.25l_x$ & 0.007316 & 0.007317 & 0.029028 & 0.029024 & 0.052692 & 0.052686 \\
		& $0.5l_x$ & 0.008052 & 0.008053 & 0.037900 & 0.037898 & 0.049495 & 0.049488 \\
		\bottomrule
		\multicolumn{8}{l}{\footnotesize Values marked with an asterisk (*) are taken from the nearest available skeleton line.}
	\end{tabular}
}
\end{table}

\subsection{Surface Blending and Boundary Matching}
\label{subsec:cosineblending}

\begin{figure}[ht]
	\centering
	\includegraphics[width=0.7\textwidth]{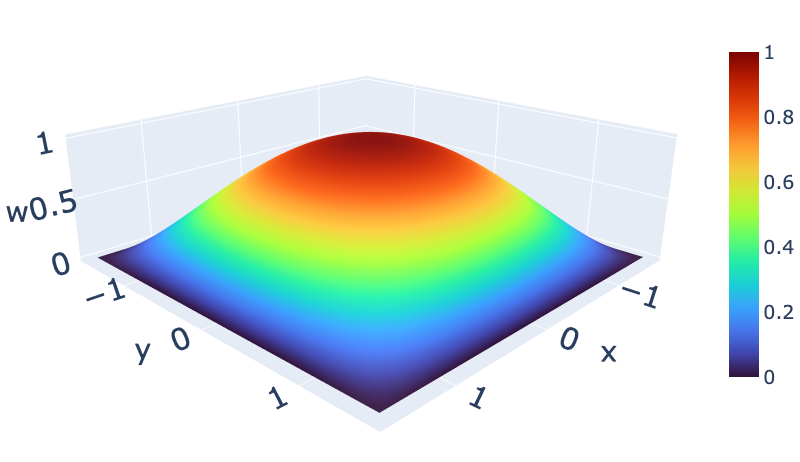}
	\caption{Cosine-like biharmonic surface patch}
	\label{fig:cosine_biharmonic}
\end{figure}

A key advantage of MMS over traditional splines is its robust handling of boundary conditions, allowing us to effectively solve blending problems and obtain the most energy-efficient surfaces. We demonstrate this by solving the blending problem for a cosine-like biharmonic function (Fig.~\ref{fig:cosine_biharmonic}):
\begin{equation}
	\begin{split}
		W_{\text{cos}}(x,y) &= \cos(x) \left( \frac{1}{2}\cosh(y) - \coth\left(\frac{\pi}{2}\right) \frac{y}{\pi}\sinh(y) \right) \\
		&+ \cos(y) \left( \frac{1}{2}\cosh(x) - \coth\left(\frac{\pi}{2}\right) \frac{x}{\pi}\sinh(x) \right)\;,
	\end{split}
	\label{eq:cosine_biharmonic_func}
\end{equation}
which is constructed to be zero at the boundaries $x=\pm\pi/2$ and $y=\pm\pi/2$, and equal to 1 at the center.

To solve this problem in MMS, we set the boundary conditions at the midpoints of the outer sides of the boundary elements. We set the elevation $w$, normal angle $\theta_n$, and tangential angle $\theta_\tau$. Elevation $w$ is set to the value of function \eqref{eq:cosine_biharmonic_func} at the boundary points. The angular boundary conditions are specified according to the first derivatives:
\begin{align}
	\theta_{n}^x &= \theta_{\tau}^y = \frac{\partial W_{\text{cos}}}{\partial x} \nonumber\\
	&= -\sin(x) \left( \frac{1}{2}\cosh(y) - \coth\left(\frac{\pi}{2}\right) \frac{y}{\pi}\sinh(y) \right) \nonumber \\
	&+ \cos(y) \left( \frac{1}{2}\sinh(x) - \coth\left(\frac{\pi}{2}\right) \frac{1}{\pi}\sinh(x) - \coth\left(\frac{\pi}{2}\right) \frac{x}{\pi}\cosh(x) \right), \label{eq:cosine_deriv_x} \\
	\theta_{n}^y &= \theta_{\tau}^x = \frac{\partial W_{\text{cos}}}{\partial y} \nonumber\\
	&= \cos(x) \left( \frac{1}{2}\sinh(y) - \coth\left(\frac{\pi}{2}\right) \frac{1}{\pi}\sinh(y) - \coth\left(\frac{\pi}{2}\right) \frac{y}{\pi}\cosh(y) \right) \nonumber \\
	&- \sin(y) \left( \frac{1}{2}\cosh(x) - \coth\left(\frac{\pi}{2}\right) \frac{x}{\pi}\sinh(x) \right). \label{eq:cosine_deriv_y}
\end{align}

Prior to assessing mesh convergence, we demonstrate the capabilities of a single MMS element. In a $1 \times 1$ mesh, the entire domain $[-\frac{\pi}{2}; \frac{\pi}{2}] \times [-\frac{\pi}{2}; \frac{\pi}{2}]$ is represented by a single rectangular element, with the displacement field reconstructed using the Coons patch formula \eqref{eq:Coons}. The orthogonal skeleton lines (the element midlines) divide the domain into four sub-patches. Each sub-patch is bounded by two skeleton line segments and two outer boundary curves, which are reconstructed via \eqref{eq:Coonsouter}. Given that these sub-patches are situated at the corners of the domain, corner displacements are determined by averaging independent extrapolations from \eqref{eq:12}, as detailed in Sect.~\ref{subsec:computing}.

\begin{figure}[ht]
	\centering
	\includegraphics[width=0.5\textwidth]{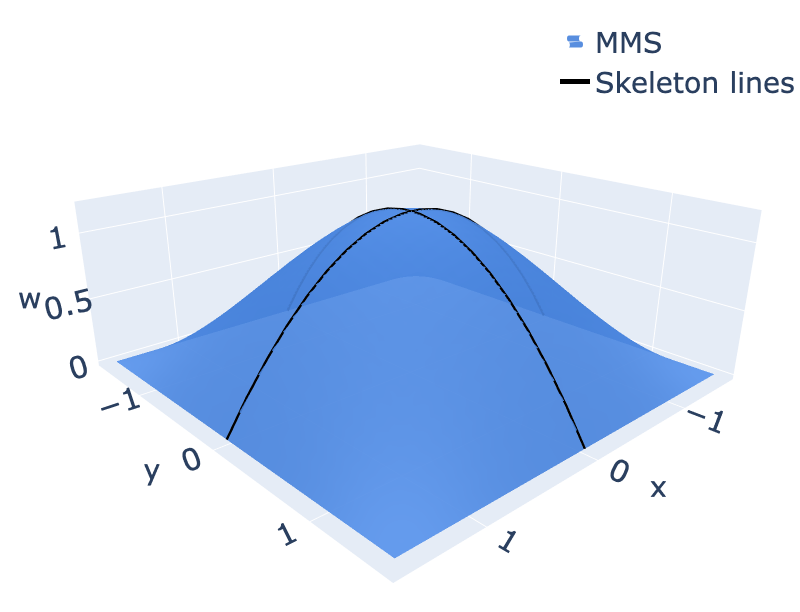}
	\caption{Cosine-like biharmonic surface reconstructed with a single MMS element ($1\times 1$ mesh, $40\times 40$ subpoints in each sub-patch)}
	\label{fig:cosine_biharmonic_single}
\end{figure}

However, since the Coons patch provides only the displacement function $W^{\text{MMS}}\left(c,d\right)$, the curvatures (moments) required for the energy functional \eqref{eq:discrete_energy_element} are not available in closed form and need to be computed numerically:
\begin{align}
	M_n^x\left(c,d\right) &\approx \frac{W^{\text{MMS}}\left(c+h, d\right)-2 W^{\text{MMS}}\left(c, d\right)+W^{\text{MMS}}\left(c-h, d\right)}{h^2}\;,\\
	M_n^y\left(c,d\right) &\approx \frac{W^{\text{MMS}}\left(c, d+h\right)-2 W^{\text{MMS}}\left(c, d\right)+W^{\text{MMS}}\left(c, d-h\right)}{h^2}\;,\\
	M_\tau\left(c,d\right) &\approx \frac{W^{\text{MMS}}\left(c+h, d+h\right)-W^{\text{MMS}}\left(c+h, d-h\right)}{4 h^2}\\
	&-\frac{W^{\text{MMS}}\left(c-h, d+h\right)+W^{\text{MMS}}\left(c-h, d-h\right)}{4 h^2}\;.
\end{align}
The energy integral \eqref{eq:discrete_energy_sum} is then evaluated via sub-grid summation of the curvature contributions \eqref{eq:discrete_energy_element}. As shown in Fig.~\ref{fig:cosine_biharmonic_single}, the reconstructed surface captures the characteristic dome-like profile of the biharmonic function \eqref{eq:cosine_biharmonic_func} despite the coarse discretization. The resulting energy functional value $E\left(W^{\text{MMS}}_{\text{cos}}\right) = 13.5406$---discussed in detail below---confirms that the patch construction procedure yields a physically meaningful and energetically consistent approximation, even when utilizing only a single MMS element.

Table \ref{tab:cosine_restore_angles} demonstrates rapid convergence. Precise results are obtained even when using only $7\times 7=49$ elements. Crucially, the reconstructed surface exhibits near-perfect energy characteristics. For a $71 \times 71$ mesh, we get $E\left(W_{\text{cos}}\right) = 17.1850$ and $E\left(W^{\text{MMS}}_{\text{cos}}\right) = 17.1785$.

\begin{table}[ht]
	\centering
	\caption{Cosine-like biharmonic surface restoration from boundary elevations and angles}
	\label{tab:cosine_restore_angles}
	\begin{tabular}{lcccccc}
		\toprule
		$N\times N$ & $w^x(0,0)$ & $\theta^x(0,0)$ & $M^x(0,0)$ & $w^x(\pi/3,0)$ & $\theta^x(\pi/3,0)$ & $M^x(\pi/3,0)$ \\
		\midrule
		7$\times$7 & 0.9973 & 0.0000 & 0.6934 & 0.5952 & 0.8213 & 1.0125 \\
		21$\times$21 & 0.9997 & 0.0000 & 0.6940 & 0.5960 & 0.8232 & 1.0156 \\
		31$\times$31 & 0.9999 & 0.0000 & 0.6941 & 0.5960 & 0.8234 & 1.0146 \\
		71$\times$71 & 1.0000 & 0.0000 & 0.6941 & 0.5961 & 0.8235 & 1.0147 \\
		Exact & 1 & 0 & 0.6941 & 0.5961 & 0.8236 & 1.0147 \\
		\bottomrule
	\end{tabular}
\end{table}

We then reconstruct the same surface using higher-order boundary data, i.e. second derivatives (curvatures) at the boundaries:
\begin{align}
	M_n^x &= \frac{\partial^2W_{\text{cos}}}{\partial x^2} = -\cos(x)\cdot\left(\frac{\cosh(y)}{2} - \coth \left(\frac{\pi}{2}\right)\cdot\frac{y\sinh(y)}{\pi}\right)\nonumber \\
	&+ \cos(y)\cdot\left(\frac{\cosh(x)}{2} - 2\coth\left(\frac{\pi}{2}\right)\cdot\frac{\cosh(x)}{\pi} - \coth\left(\frac{\pi}{2}\right)\cdot\frac{x\sinh(x)}{\pi}\right),\\
	M_n^y &= \frac{\partial^2W_{\text{cos}}}{\partial y^2} = - \cos(y)\cdot\left(\frac{\cosh(x)}{2} - \coth\left(\frac{\pi}{2}\right)\cdot\frac{x\sinh(x)}{\pi}\right)\nonumber \\
	&+  \cos(x)\cdot\left(\frac{\cosh(y)}{2} - 2\coth \left(\frac{\pi}{2}\right)\cdot\frac{\cosh(y)}{\pi} -\coth\left(\frac{\pi}{2}\right)\cdot\frac{y\sinh(y)}{\pi}\right)\;.
\end{align}
Instead of setting tangential angles, we set the values for tangential curvatures:
\begin{align}
	M_{\tau}^x &= M_{\tau}^y = \frac{\partial^2 W_{\text{cos}}}{\partial x \partial y} \nonumber\\
	&= -\sin(x) \cdot \left(\frac{\sinh(y)}{2} - \coth\left(\frac{\pi}{2}\right)\cdot \frac{\sinh(y)}{\pi}  - \coth\left(\frac{\pi}{2}\right)\cdot\frac{y\cosh(y)}{\pi}\right)\nonumber \\
	&- \sin(y)\cdot \left(\frac{\sinh(x)}{2} - \coth\left(\frac{\pi}{2}\right)\cdot\frac{\sinh(x)}{\pi} - \coth\left(\frac{\pi}{2}\right)\cdot\frac{x\cosh(x)}{\pi}\right)\;.
\end{align}

The comparison results in Table \ref{tab:cosine_restore_moments} show accurate convergence even for these specific boundary conditions. The energy functional for the $71 \times 71$ grid is $E\left(W^{\text{MMS}}_{\text{cos}}\right) = 17.1821$, which corresponds to an extremely smooth bi-harmonic function. Notably, the energy value derived from the single-element Coons patch, equal to $13.5406$, is lower than the exact solution. This discrepancy arises because the Coons patch lacks $C^2$ continuity across sub-patch boundaries; consequently, the energy functional cannot be validly evaluated for this approximation.

\begin{table}[ht!]
	\centering
	\caption{Cosine-like biharmonic surface restoration from boundary elevations and second derivatives}
	\label{tab:cosine_restore_moments}
	\begin{tabular}{lcccccc}
		\toprule
		$N\times N$ & $w^x(0,0)$ & $\theta^x(0,0)$ & $M^x(0,0)$ & $w^x(\pi/3,0)$ & $\theta^x(\pi/3,0)$ & $M^x(\pi/3,0)$ \\
		\midrule
		7$\times$7 & 1.0062 & 0.0000 & 0.6996 & 0.6004 & 0.8288 & 1.0211 \\
		21$\times$21 & 1.0008 & 0.0000 & 0.6948 & 0.5966 & 0.8242 & 1.0167 \\
		31$\times$31 & 1.0004 & 0.0000 & 0.6944 & 0.5963 & 0.8238 & 1.0152 \\
		71$\times$71 & 1.0001 & 0.0000 & 0.6942 & 0.5961 & 0.8236 & 1.0148 \\
		Exact & 1 & 0 & 0.6941 & 0.5961 & 0.8236 & 1.0147 \\
		\bottomrule
	\end{tabular}
\end{table}

\begin{figure}[ht!]
	\centering
	\includegraphics[width=0.5\textwidth]{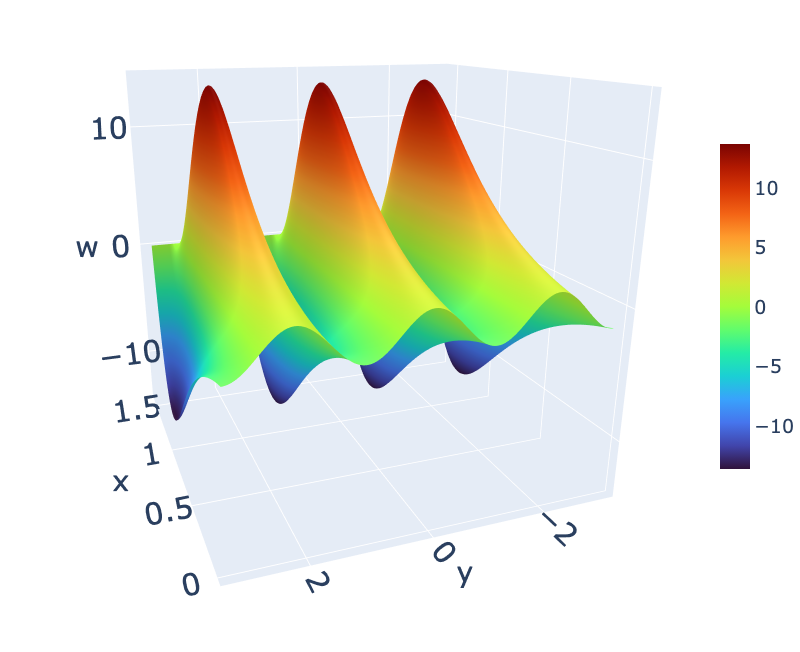}
	\caption{Non-symmetric biharmonic surface}
	\label{fig:nonsymm_biharmonic}
\end{figure}

We now consider a more complex, asymmetrical function in the domain $[0; \pi/2] \times [-\pi; \pi]$ (Fig.~\ref{fig:nonsymm_biharmonic}):
\begin{equation}
	W_{\text{asy}}(x,y) = \left(\frac{\pi}{2} - x\right) e^{3x} \cos(3y)\;.
	\label{eq:nonsymm_func}
\end{equation}

By analogy with the previous example, we set the values of the function and its first derivatives at the boundaries. The first derivatives are:
\begin{align}
	\theta_{n}^x = \theta_{\tau}^y = \frac{\partial W_{\text{asy}}}{\partial x} &= \left(\frac{3\pi}{2} - 3x - 1\right) e^{3x} \cos(3y)\;, \label{eq:nonsymm_deriv_x} \\
	\theta_{n}^y = \theta_{\tau}^x = \frac{\partial W_{\text{asy}}}{\partial y} &= -3\left(\frac{\pi}{2} - x\right) e^{3x} \sin(3y)\;. \label{eq:nonsymm_deriv_y}
\end{align}
Table \ref{tab:nonsymm_restore_angles} shows that a larger number of elements is required to obtain high-precision results, but a smooth surface can be constructed even with the smallest mesh. For a $251 \times 251$ grid, the energy functionals are $E(W_{\text{asy}}) = 233228.4505$ and $E\left(W^{\text{MMS}}_{\text{asy}}\right) = 233201.0544$.

\begin{table}[ht!]
	\centering
	\caption{Restoration of biharmonic function \eqref{eq:nonsymm_func} from its boundary elevations and angles}
	\label{tab:nonsymm_restore_angles}
	\resizebox{\textwidth}{!}{%
		\begin{tabular}{lcccccc}
			\toprule
			$N\times N$ & $w^x(\pi/4,0)$ & $\theta^x(\pi/4,0)$ & $M^x(\pi/4,0)$ & $w^x(3\pi/8,0)$ & $\theta^x(3\pi/8,0)$ & $M^x(3\pi/8,0)$ \\
			\midrule
			21$\times$21 & 8.1251 & 14.1549 & 11.9868 & 13.3142 & 6.3979 & 83.1234 \\
			51$\times$51 & 8.2612 & 14.2841 & 11.3837 & 13.4350 & 6.1488 & 84.2887 \\
			101$\times$101 & 8.2801 & 14.3026 & 11.3018 & 13.4523 & 6.1148 & 84.4478 \\
			251$\times$251 & 8.2855 & 14.3078 & 11.2788 & 13.4572 & 6.1053 & 84.4930 \\
			Exact & 8.2865 & 14.3088 & 11.2743 & 13.4581 & 6.1035 & 84.5016 \\
			\bottomrule
		\end{tabular}
	}
\end{table}

Specifying the boundaries with elevation and second derivatives (Table~\ref{tab:nonsymm_restore_moments}) yields an energy functional of $E\left(W^{\text{MMS}}_{\text{asy}}\right) = 233192.9470$ for the $251 \times 251$ grid. Thus, MMS allows for extremely flexible ways to specify boundary conditions and effectively solves the problem of blending with smooth surfaces.

\begin{table}[ht!]
	\centering
	\caption{Restoration of biharmonic function \eqref{eq:nonsymm_func} from its boundary elevations and curvatures}
	\label{tab:nonsymm_restore_moments}
	\resizebox{\textwidth}{!}{%
		\begin{tabular}{lcccccc}
			\toprule
			$N\times N$ & $w^x(\pi/4,0)$ & $\theta^x(\pi/4,0)$ & $M^x(\pi/4,0)$ & $w^x(3\pi/8,0)$ & $\theta^x(3\pi/8,0)$ & $M^x(3\pi/8,0)$ \\
			\midrule
			21$\times$21 & 8.0376 & 14.0985 & 11.9612 & 13.2100 & 6.4043 & 82.5687 \\
			51$\times$51 & 8.2485 & 14.2773 & 11.3828 & 13.4206 & 6.1511 & 84.2140 \\
			101$\times$101 & 8.2770 & 14.3010 & 11.3017 & 13.4488 & 6.1155 & 84.4296 \\
			251$\times$251 & 8.2850 & 14.3076 & 11.2788 & 13.4566 & 6.1055 & 84.4901 \\
			Exact & 8.2865 & 14.3088 & 11.2743 & 13.4581 & 6.1035 & 84.5016 \\
			\bottomrule
		\end{tabular}
	}
\end{table}

\subsection{Interpolation From Sparse Internal Data}

\begin{figure}[ht!]
	\centering
	\includegraphics[width=0.5\textwidth]{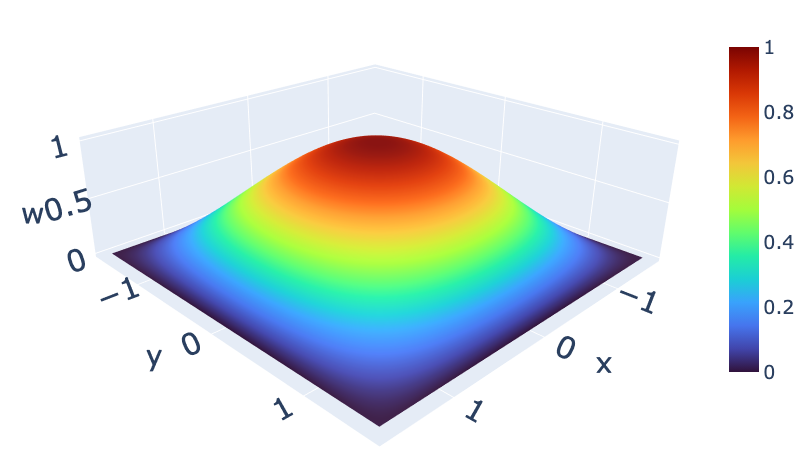}
	\caption{Cosine-based surface}
	\label{fig:cosine_based_surface}
\end{figure}

We now address surface reconstruction from sparse internal data points (interpolation).
Let us consider constructing surfaces based on boundary as well as internal input points. We will demonstrate the smoothness and precision of the resulting splines by restoring a surface generated by a twice continuously differentiable, but non-biharmonic, function in the interval $x,y \in [-\pi/2; \pi/2]$ (Fig.~\ref{fig:cosine_based_surface}):
\begin{equation}
	W_{\text{cos2}}(x,y) = \cos(x) \cdot \cos(y).
	\label{eq:cosine_based_func}
\end{equation}
The energy of this surface is $E(W_{\text{cos2}}) = \pi^2 \approx 9.8696$.

\begin{figure}[ht!]
	\centering
	\begin{subfigure}[b]{0.48\textwidth}
		\centering
		\includegraphics[width=\textwidth]{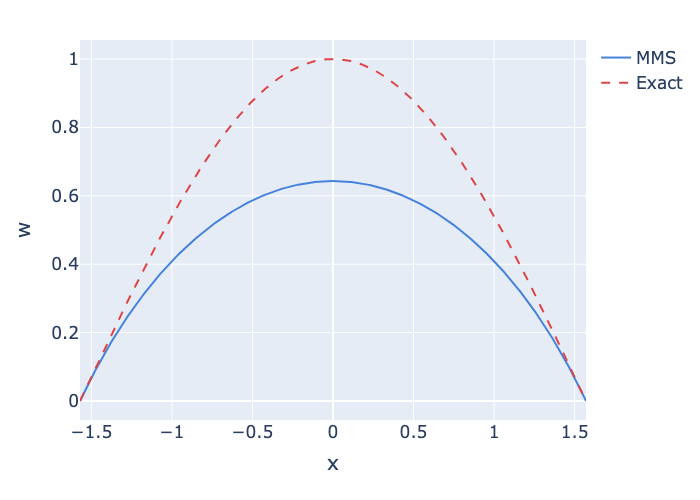}
		\caption{Surface elevation}
	\end{subfigure}
	\hfill
	\begin{subfigure}[b]{0.48\textwidth}
		\centering
		\includegraphics[width=\textwidth]{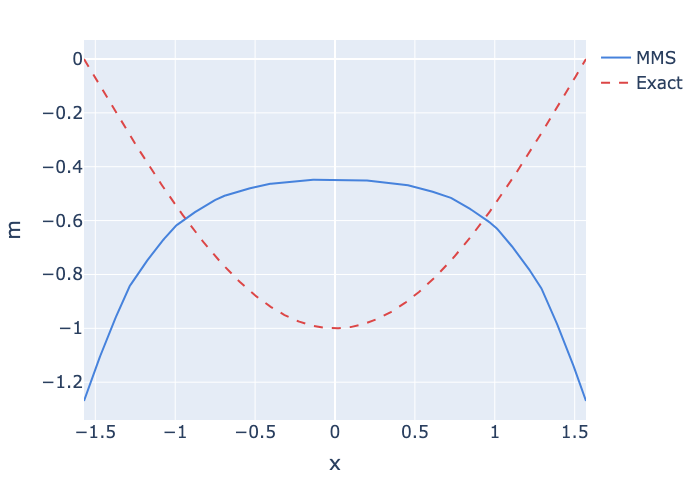}
		\caption{Curvatures}
	\end{subfigure}
	\caption{Cosine-based results along the line $y=0$ ($11 \times 11$, no inputs)}
	\label{fig:cosine_results_no_input}
\end{figure}

First, we restore the surface using only boundary conditions (elevations and first derivatives), like in the previous examples. In particular, the first derivatives are:
\begin{align*}
	\theta_n^x &= \theta_\tau^y = \frac{\partial W_{\text{cos2}}}{\partial x} = -\sin(x)\cdot \cos(y)\;,\\
	\theta_n^y &= \theta_\tau^x = \frac{\partial W_{\text{cos2}}}{\partial y} = -\cos(x)\cdot \sin(y)\;.
\end{align*}
For a small $11 \times 11$ grid, the maximum elevation is 0.6431, far from the original's central value of 1 (Fig.~\ref{fig:cosine_results_no_input}). However, its energy, $E\left(W^{\text{MMS}}_{\text{cos2}}\right) = 6.9089$, is significantly lower (better) than the original surface's energy. The energy obtained for MMS in this case has an additional error due to averaging over large elements. Unlike a cosine-based function, the surface constructed by MMS is bi-harmonic and therefore has the optimal energy characteristics given the boundary conditions. However, the surfaces do not match well.

It should be noted that the energy degraded not only because of the stronger bend in the center, but also because any deviation from MMS (which satisfies the biharmonic equation) will ultimately result in a lower quality surface. Substituting the condition $W^{\text{MMS}}_{\text{cos2}}(0, 0)=0.3$ (lower than for the biharmonic functions, i.e. 0.6431), we still get worse energy:  $E\left(W^{\text{MMS}}_{\text{cos2}}\right) = 8.9159$.

\begin{figure}[ht!]
	\centering
	\begin{subfigure}[b]{0.48\textwidth}
		\centering
		\includegraphics[width=\textwidth]{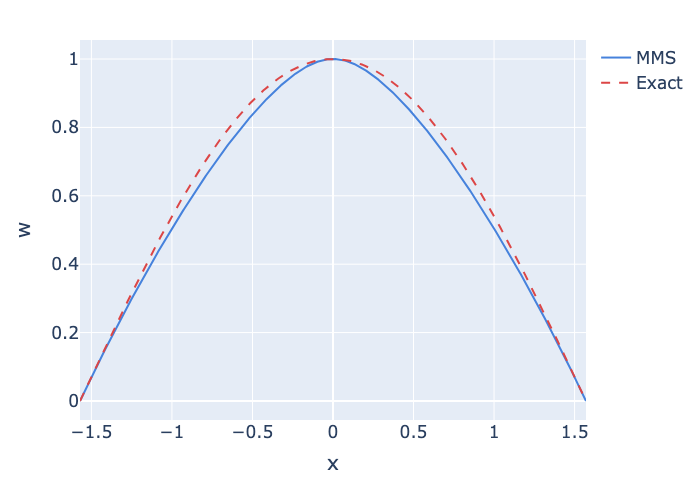}
		\caption{Surface elevation}
	\end{subfigure}
	\hfill
	\begin{subfigure}[b]{0.48\textwidth}
		\centering
		\includegraphics[width=\textwidth]{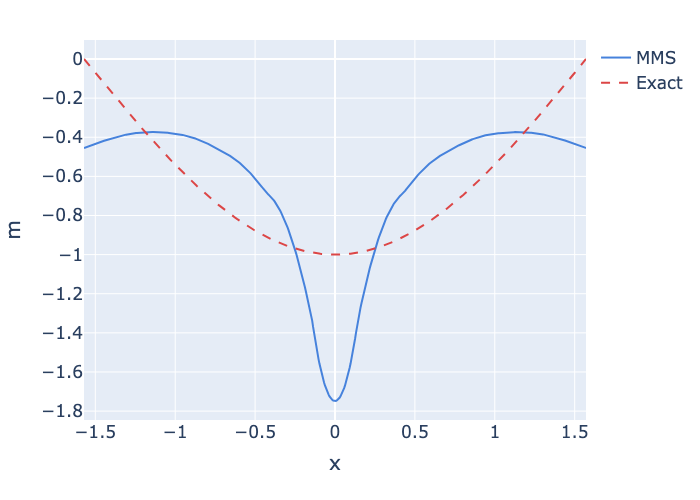}
		\caption{Curvatures}
	\end{subfigure}
	\caption{Cosine-based results along the line $y=0$ ($11 \times 11$, central input)}
	\label{fig:cosine_results_central_input}
\end{figure}

\begin{figure}[ht!]
	\centering
	\begin{subfigure}[b]{0.48\textwidth}
		\centering
		\includegraphics[width=\textwidth]{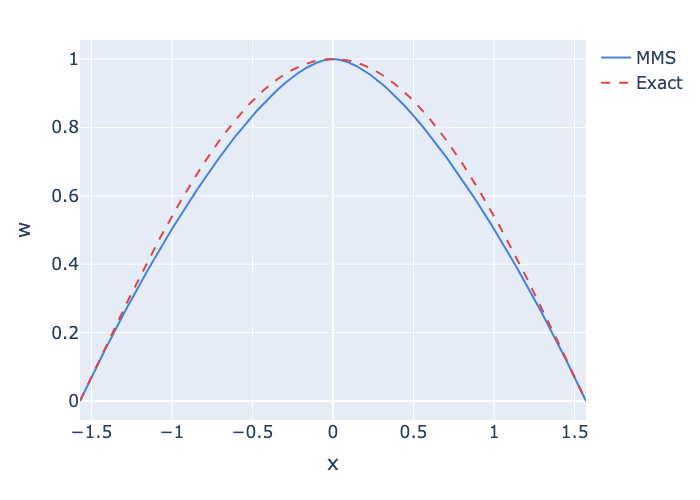}
		\caption{Surface elevation}
	\end{subfigure}
	\hfill
	\begin{subfigure}[b]{0.48\textwidth}
		\centering
		\includegraphics[width=\textwidth]{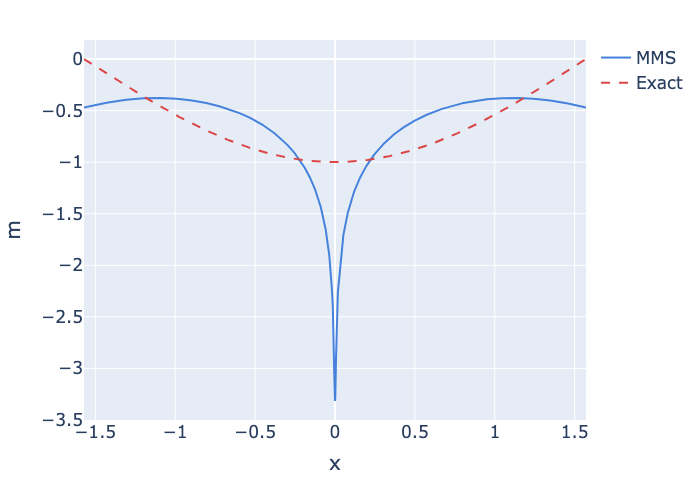}
		\caption{Curvatures}
	\end{subfigure}
	\caption{Cosine-based results along the line $y=0$ ($251 \times 251$, central input)}
	\label{fig:cosine_results_fine_mesh}
\end{figure}

To match the original surface, we introduce a constraint at the central point, $W^{\text{MMS}}_{\text{cos2}}(0,0)=1$. The elevation now matches the original, but the energy has increased to $E\left(W^{\text{MMS}}_{\text{cos2}} \right) = 9.3161$ (Fig.~\ref{fig:cosine_results_central_input}). Although the obtained surface is quite close to the original, certain calculations require a much denser mesh.  On a finer $251 \times 251$ grid, a third derivative discontinuity develops at the central point (Fig.~\ref{fig:cosine_results_fine_mesh}). Indeed, constraining $w$ at the center engenders third derivative $Q$ in this element, which introduces a large gradient of the third derivative between the constrained element and its unconstrained neighbors.  This behavior corresponds to the general property of the thin plate particular solution, where the second derivative contains a singularity.

In our case, the concentrated force is spreading over the element, so the larger the element, the smaller the peak of the second derivative. Despite this, the total surface energy remains at the same level $E\left(W^{\text{MMS}}_{\text{cos2}}\right) = 9.3145$, which is still better than the original surface energy.

\begin{figure}[ht!]
	\centering
	\begin{subfigure}[b]{0.48\textwidth}
		\centering
		\includegraphics[width=\textwidth]{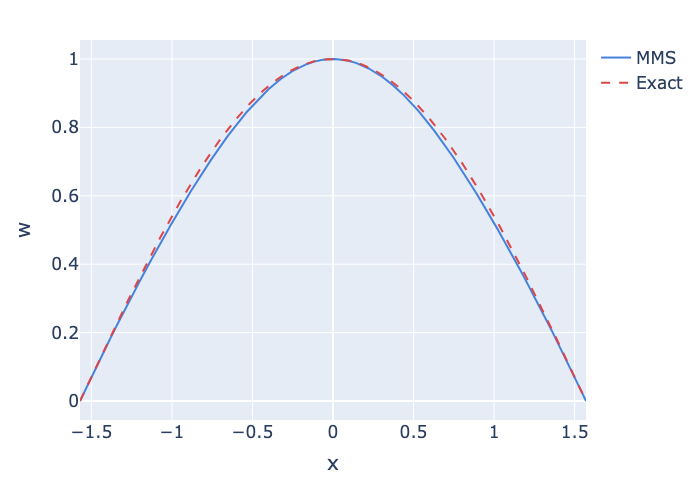}
		\caption{Surface elevation}
	\end{subfigure}
	\hfill
	\begin{subfigure}[b]{0.48\textwidth}
		\centering
		\includegraphics[width=\textwidth]{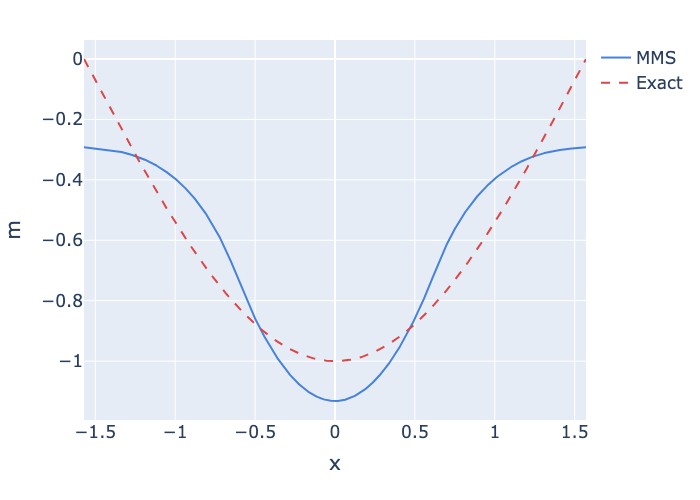}
		\caption{Curvatures}
	\end{subfigure}
	\caption{Cosine-based results along the line $y=0$ ($251 \times 251$, central input, $\zeta=50$)}
	\label{fig:cosine_results_spreading}
\end{figure}

As discussed in Sect.~\ref{subsec:reg}, constraining a single point induces a third-derivative discontinuity (Fig.~\ref{fig:cosine_results_fine_mesh}). To mitigate this and produce a visually fair surface, we apply the regularization kernel defined in \eqref{eq:force_spreading}. For this grid, setting the regularization parameter to $\zeta=50$ significantly improves the visual smoothness while keeping the total surface energy practically unchanged at $E\left(W^{\text{MMS}}_{\text{cos2}}\right) = 9.5280$ (Fig.~\ref{fig:cosine_results_spreading}).

\begin{figure}[ht!]
	\centering
	\begin{subfigure}[b]{0.48\textwidth}
		\centering
		\includegraphics[width=\textwidth]{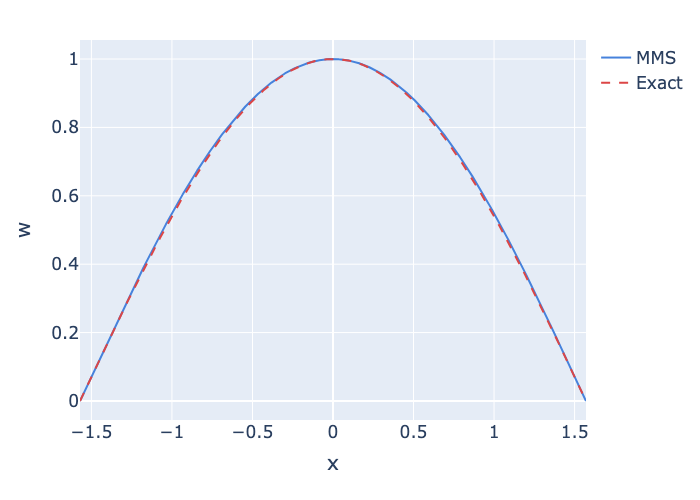}
		\caption{Surface elevation}
	\end{subfigure}
	\hfill
	\begin{subfigure}[b]{0.48\textwidth}
		\centering
		\includegraphics[width=\textwidth]{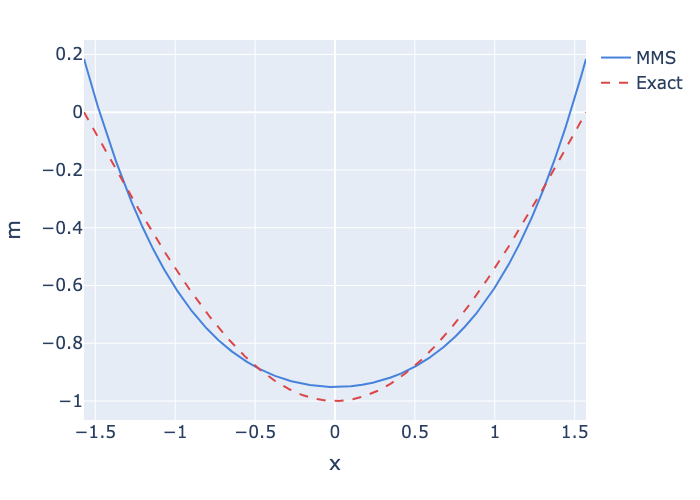}
		\caption{Curvatures}
	\end{subfigure}
	\caption{Cosine-based results along the line $y=0$ ($251 \times 251$, central input, $\zeta=10^6$)}
	\label{fig:cosine_results_spreading2}
\end{figure}

We can also exclude regularization by either setting the second part of the denominator \eqref{eq:force_spreading} to zero or assuming $\zeta$ to be very large. Doing so will give us the best approximation for the case of central input (Fig.~\ref{fig:cosine_results_spreading2}).

\begin{figure}[ht!]
	\centering
	\begin{subfigure}[b]{0.48\textwidth}
		\centering
		\includegraphics[width=\textwidth]{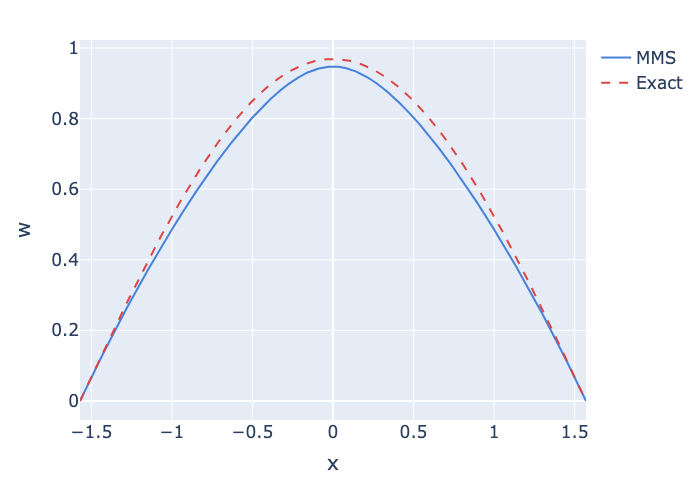}
		\caption{Surface elevation}
	\end{subfigure}
	\hfill
	\begin{subfigure}[b]{0.48\textwidth}
		\centering
		\includegraphics[width=\textwidth]{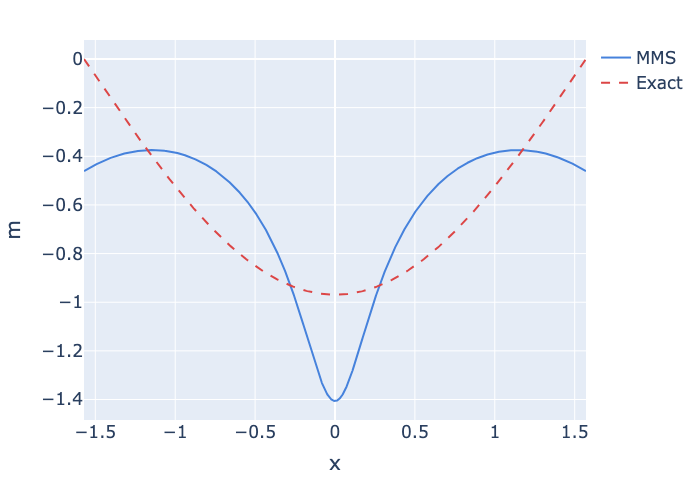}
		\caption{Curvatures}
	\end{subfigure}
	\caption{Cosine-based results along the line $y=0.25$ ($251 \times 251$, central input, without regularization)}
	\label{fig:cosine_results_spreading3}
\end{figure}

We also demonstrate that the third derivative discontinuity arises only in the vicinity of the attachment point. Returning to the configuration shown in Fig.~\ref{fig:cosine_results_fine_mesh},  if the bending curvatures are taken along an off-center skeleton line $y=0.25$ instead of the central line $y=0$, the distribution becomes considerably smoother even without the regularization technique, as shown in Fig.~\ref{fig:cosine_results_spreading3}.

\begin{figure}[ht!]
	\centering
	\begin{subfigure}[b]{0.48\textwidth}
		\centering
		\includegraphics[width=\textwidth]{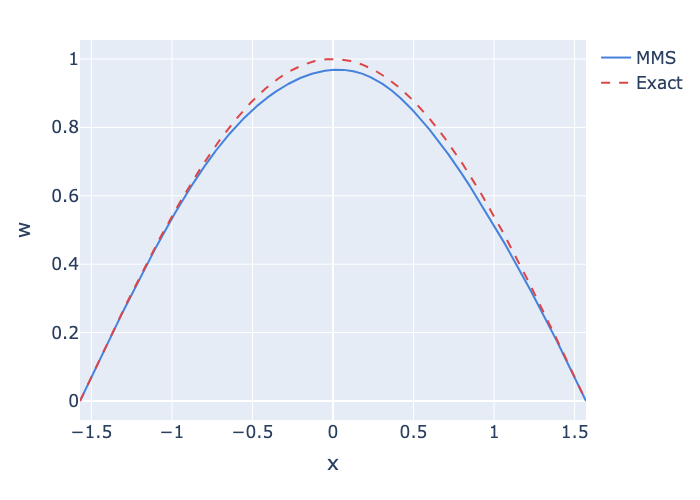}
		\caption{Surface elevation}
	\end{subfigure}
	\hfill
	\begin{subfigure}[b]{0.48\textwidth}
		\centering
		\includegraphics[width=\textwidth]{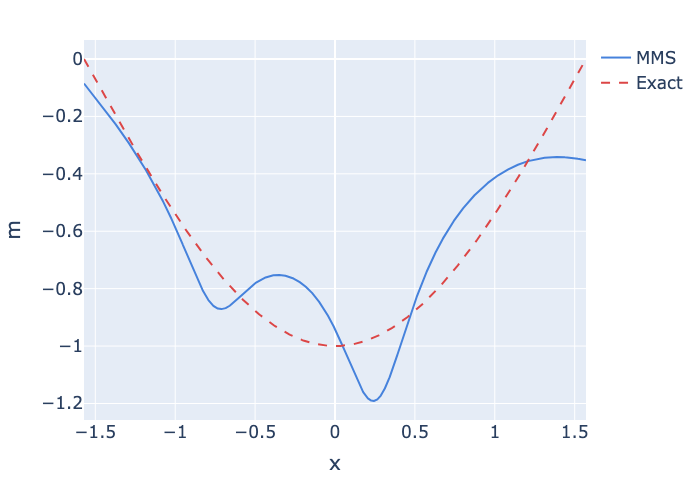}
		\caption{Curvatures}
	\end{subfigure}
	\caption{Cosine-based results along the line $y=0$ ($251 \times 251$, 5 inputs, without regularization)}
	\label{fig:cosine_results_5_inputs}
\end{figure}

Another important advantage of MMS is its ability to operate with an irregular set of input points. Suppose we are given 5 input points chosen randomly; for convenience, we will align them with the centers of certain elements:
\begin{equation}
\left(\frac{20\pi}{251},\frac{20\pi}{251}\right);\left(\frac{40\pi}{251},\frac{80\pi}{251}\right);\left(-\frac{90\pi}{251},\frac{70\pi}{251}\right);\left(-\frac{30\pi}{251},-\frac{80\pi}{251}\right);\left(-\frac{60\pi}{251},-\frac{20\pi}{251}\right)
\label{eq:pis}
\end{equation}
From the results shown in Fig.~\ref{fig:cosine_results_5_inputs}, even with such an asymmetric set of points, the restoring quality remains very high, while the surface energy still outperforms that of the original surface: $E\left(W^{\text{MMS}}_{\text{cos2}}\right) = 9.5445$.

\begin{figure}[ht!]
	\centering
	\begin{subfigure}[b]{0.48\textwidth}
		\centering
		\includegraphics[width=\textwidth]{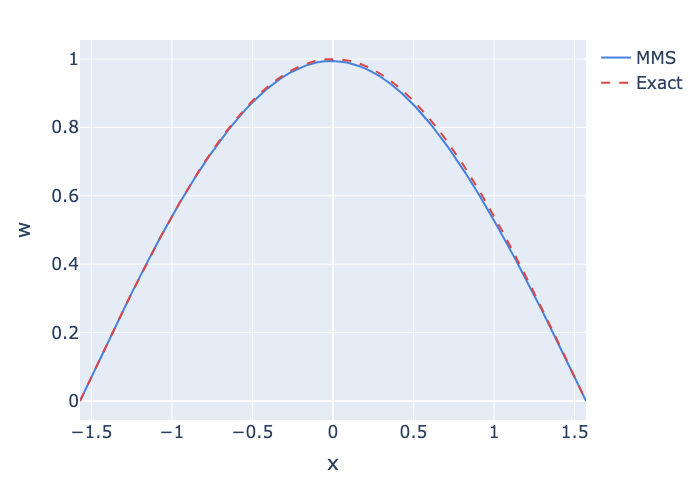}
		\caption{Surface elevation}
	\end{subfigure}
	\hfill
	\begin{subfigure}[b]{0.48\textwidth}
		\centering
		\includegraphics[width=\textwidth]{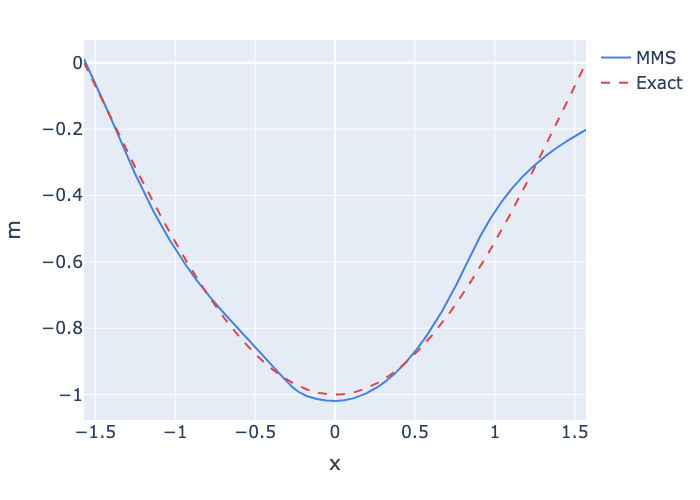}
		\caption{Curvatures}
	\end{subfigure}
	\caption{Cosine-based results along the line $y=0$ ($251 \times 251$, 5 input, $\zeta=50$)}
	\label{fig:cosine_results_5_inputs2}
\end{figure}

We can also ``spread'' the load from the elements containing the input points (Fig.~\ref{fig:cosine_results_5_inputs2}). The resulting energy is $E\left(W^{\text{MMS}}_{\text{cos2}}\right) = 9.7271$.

\begin{figure}[ht!]
	\centering
	\includegraphics[width=0.7\textwidth]{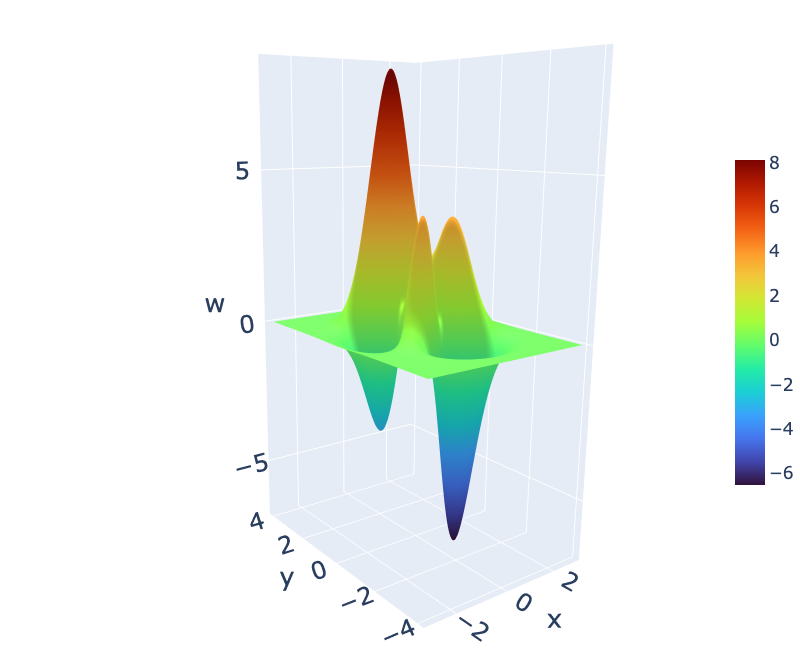}
	\caption{Multipeak surface}
	\label{fig:multipeak_surface}
\end{figure}

Let us consider a much more complex surface, devoid of symmetries, taken from \cite{Yi2021} and defined in the region $[-3; 3] \times [-4; 4]$ (Fig.~\ref{fig:multipeak_surface}):
\begin{equation}
	\begin{split}
		W_{Yi}(x,y) &= 3(1-x)^2 e^{-x^2-(y+1)^2} - 10\left(\frac{1}{5}x-x^3-y^5\right)e^{-x^2-y^2} \\
		&- \frac{1}{3}e^{-(x+1)^2-y^2}\;.
	\end{split}
	\label{eq:multipeak_func}
\end{equation}
It has three local minima and three local maxima near $(0, 0)$, and at the boundaries, the function and its derivatives vanish. The energy functional for this surface is $E(W_{Yi}) = 4161.9368$.

\begin{figure}[ht!]
	\centering
	\begin{subfigure}[b]{0.48\textwidth}
		\centering
		\includegraphics[width=\textwidth]{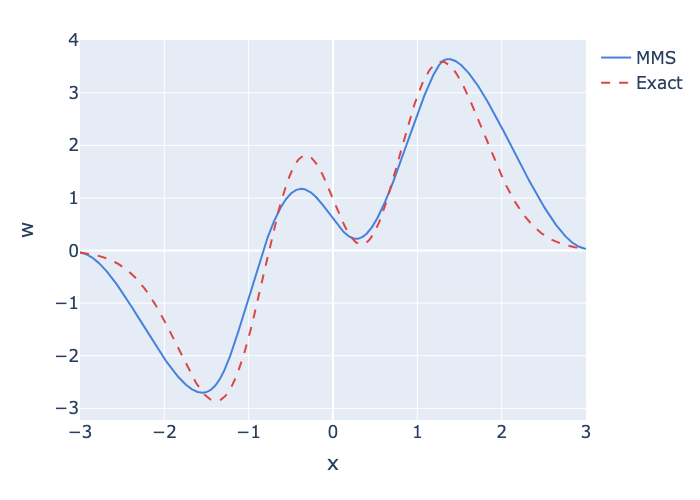}
		\caption{Surface elevation}
	\end{subfigure}
	\hfill
	\begin{subfigure}[b]{0.48\textwidth}
		\centering
		\includegraphics[width=\textwidth]{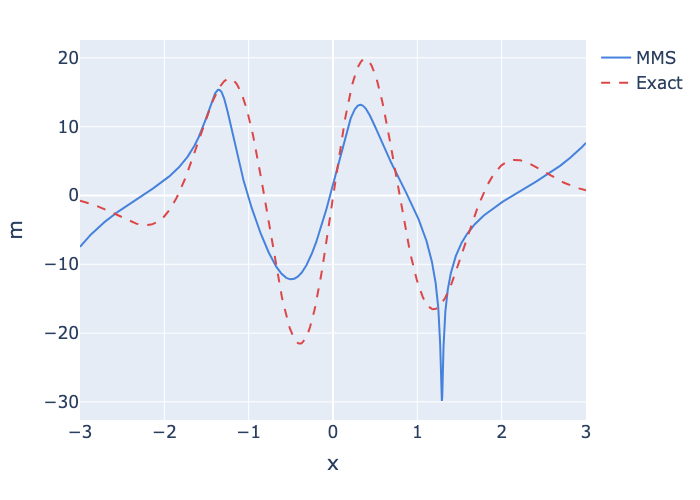}
		\caption{Curvatures}
	\end{subfigure}
	\caption{Multipeak results along the line $y=0$ ($251 \times 251$, 6 inputs)}
	\label{fig:multipeak_results_6_inputs}
\end{figure}

\begin{figure}[ht!]
	\centering
	\begin{subfigure}[b]{0.48\textwidth}
		\centering
		\includegraphics[width=\textwidth]{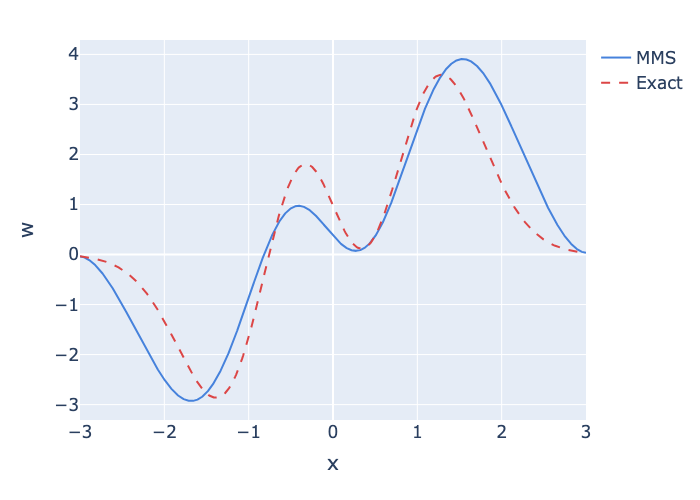}
		\caption{Surface elevation}
	\end{subfigure}
	\hfill
	\begin{subfigure}[b]{0.48\textwidth}
		\centering
		\includegraphics[width=\textwidth]{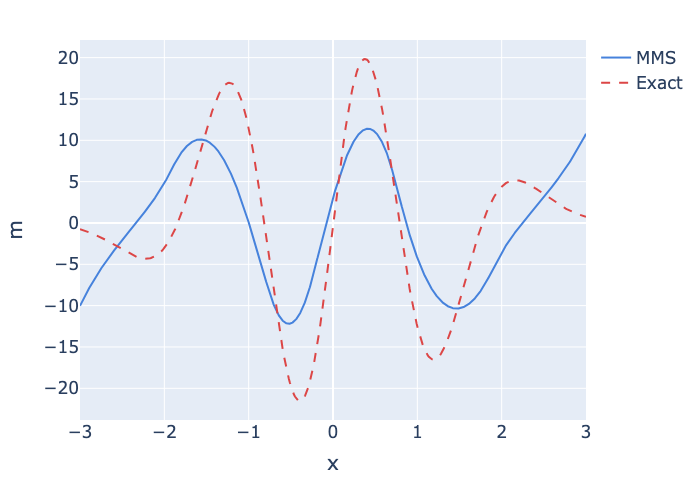}
		\caption{Curvatures}
	\end{subfigure}
	\caption{Multipeak results along the line $y=0$ ($251 \times 251$, 6 inputs, $\zeta=50$)}
	\label{fig:multipeak_results_6_inputs_spreading}
\end{figure}

Using a $251 \times 251$ grid and 6 entry points at the local extrema, we can approximate the original surface quite accurately (Fig.~\ref{fig:multipeak_results_6_inputs}), with the approximate coordinates being $(0, 1.59)$, $(-0.45, -0.64)$, $(1.29, 0)$, $(0.24, -1.63)$, $(-1.34, 0.19)$, $(0.29, 0.32)$. From an energy perspective, the resulting surface is of much higher quality, $E\left(W^{\text{MMS}}_{Yi}\right) = 2731.5228$. Applying the regularization technique significantly smooths the third derivatives near the attachment points, though it increases the energy to $E\left(W^{\text{MMS}}_{Yi}\right) = 3112.1235$ (Fig.~\ref{fig:multipeak_results_6_inputs_spreading}).

\begin{figure}[ht!]
	\centering
	\begin{subfigure}[b]{0.48\textwidth}
		\centering
		\includegraphics[width=\textwidth]{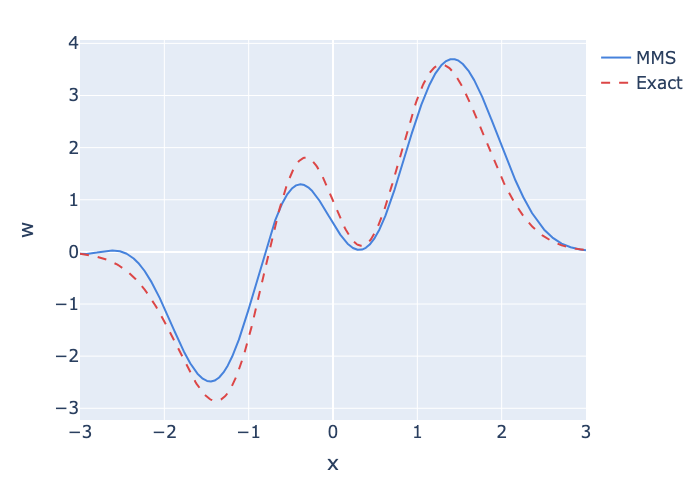}
		\caption{Surface elevation}
	\end{subfigure}
	\hfill
	\begin{subfigure}[b]{0.48\textwidth}
		\centering
		\includegraphics[width=\textwidth]{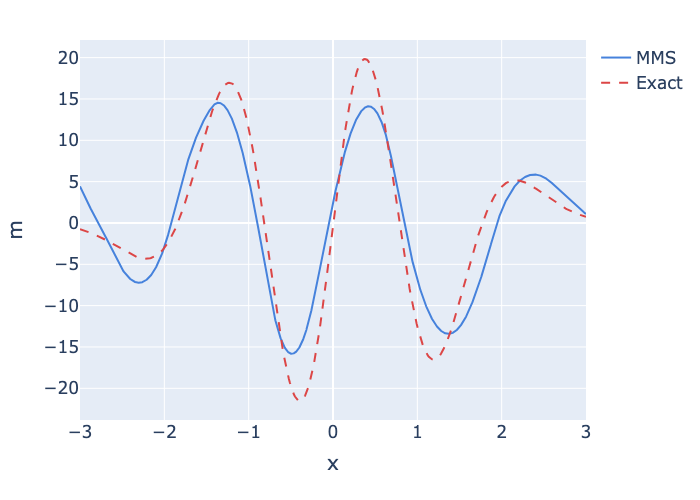}
		\caption{Curvatures}
	\end{subfigure}
	\caption{Multipeak results along the line $y=0$ ($251 \times 251$, 15 inputs, $\zeta=50$)}
	\label{fig:multipeak_results_15_inputs}
\end{figure}

Adding 9 more random points (for a total of 15) with regularization significantly improves the reconstruction accuracy, while the surface energy approaches the original: $E\left(W^{\text{MMS}}_{Yi}\right) = 3920.1770$ (Fig.~\ref{fig:multipeak_results_15_inputs}).

\begin{figure}[ht!]
	\centering
	\includegraphics[width=0.5\textwidth]{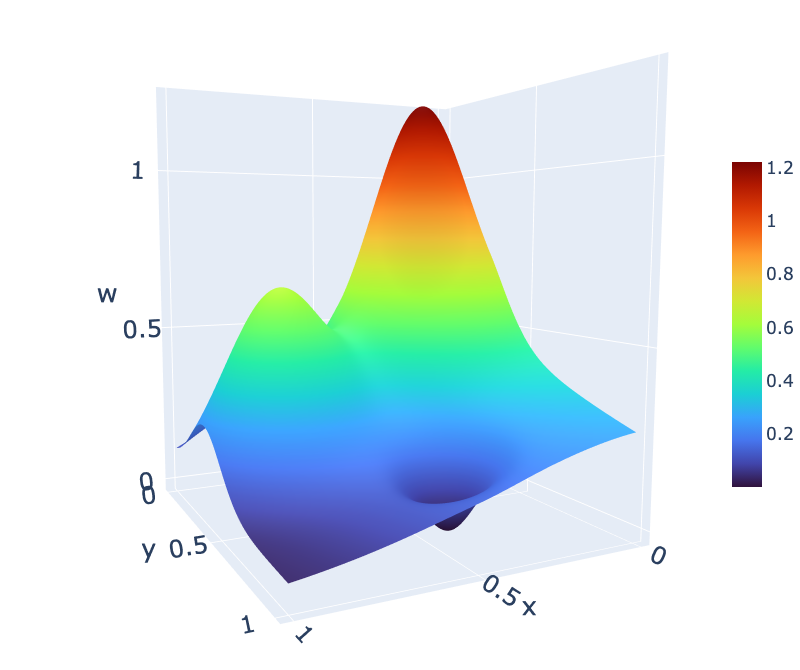}
	\caption{Franke's function in the domain $[0; 1] \times [0, 1]$}
	\label{fig:franke}
\end{figure}

Finally, let us consider the reconstruction of a well-known benchmark in the scattered data fitting literature---the Franke's function \cite{Franke}:
\begin{equation}
	\begin{split}
		W_{\text{Franke}}\left(x,y\right) &= 0.75 \exp\left(-0.25\left(9x-2\right)^2 - 0.25\left(9y - 2\right)^2\right) \\
		&+ 0.75\exp\left(-\frac{\left(9x+1\right)^2}{49} - \frac{\left(9y+1\right)}{10}\right) \\
		&+ 0.5\exp\left(-0.25\left(9x-7\right)^2 - 0.25\left(9y-3\right)^2\right) \\
		&- 0.2\exp\left(-\left(9x-4\right)^2-\left(9y-7\right)^2\right)\;,
	\end{split}
	\label{eq:franke}
\end{equation}
limited to domain $[0; 1] \times [0, 1]$ (Fig.~\ref{fig:franke}). Unlike the multipeak function \eqref{eq:multipeak_func}, Franke's function has non-trivial values and gradients along all four edges of the domain, so the choice of boundary conditions strongly affects the reconstruction. The energy of this surface is $E(W_{\text{Franke}}) = 214.1254$. 

We use a $251\times 251$ grid and investigate two regimes of boundary conditions combined with varying numbers of internal input points:
\begin{enumerate}
	\item In the first regime, we set the displacement $w$, normal angle $\theta_n$, and tangent angle $\theta_\tau$ using the values and first derivatives of \eqref{eq:franke} at the boundary midpoints, line in to Section~\ref{subsec:cosineblending}. 
	
	\item In the second regime, we impose force-free boundary conditions on all four sides: $Q^x = M^x_n = M^x_\tau = 0$ for horizontal beams and $Q^y = M^y_n = M^y_\tau = 0$ for vertical beams. In this case, no information about the target surface enters through the boundaries, and the reconstruction relies entirely on the internal input points.
\end{enumerate}

For both regimes, input points are snapped to the centers of an underlying $251 \times 251$ element grid using two distribution strategies:
\begin{itemize}
	\item regular grids: points are arranged in $k \times k$ lattices, $k \in \{3, 7, 11, 25\}$. Indices are selected to be approximately equispaced along each axis, with input points situated at the centers of the resulting grid elements;
	
	\item random sampling: $N$ distinct points, $N \in \{10, 25, 50, 100\}$, are selected by drawing element indices from a discrete uniform distribution without replacement. A fixed random seed is employed to ensure reproducibility.
\end{itemize}
The displacement values in both regimes at the input points are taken from \eqref{eq:franke}, and the regularization technique \eqref{eq:force_spreading} with $\zeta=50$ is applied in all cases.

In addition to the energy functional \eqref{eq:energy_functional}, it is instructive to evaluate the pointwise fidelity of the reconstructed surface relative to the target function. To this end, we define the squared displacement error as
\begin{equation}
	I(W, W^\prime) = \iint_\Omega \left(W(x, y) - W^\prime(x, y)\right)^2 dxdy\;.
	\label{eq:L2}
\end{equation}
For Franke's function, we evaluate $I(W^{\text{MMS}}_{\text{Franke}}, W_{\text{Franke}})$ by summation on the same $251\times 251$ grid as the energy. The integral of the squared target itself is
\begin{equation}
	I(W_{\text{Franke}}) = \iint_\Omega \left(W_{\text{Franke}}^2(x, y)\right)^2 dxdy = 0.2482\;.
	\label{eq:L2franke}
\end{equation}

While $E(W_{\text{Franke}})$ characterizes the smoothness (curvature content) of the surface, $I\left(W^{\text{MMS}}_{\text{Franke}}, W_{\text{Franke}}\right)$ characterizes how far the reconstructed surface departs from the target in absolute height. Together they allow us to distinguish a surface that is energetically optimal from the one that is geometrically close to a given function.

The results are given in Table~\ref{table:franke}. As expected, in all cases the MMS energy stays below $E(W_{\text{Franke}})$ and approaches it monotonically as the number of input points grows. With correct boundary conditions, already a small number of points provides a reasonable approximation, since the boundary itself carries substantial information about the target surface. With free boundary conditions, the reconstruction is much more challenging: the $3\times 3$ regular grid yields a nearly flat surface, with $E\left(W^{\text{MMS}}_{\text{Franke}}\right) = 0.63$, but the method recovers quickly as more points are added. At the $25\times 25$ regular grid, both regimes practically coincide with the exact value, confirming that the influence of the boundary conditions vanishes as the internal data becomes dense.

\begin{table}[h]
	\centering
	\caption{Reconstruction of Franke's function with correct and free boundary conditions for varying numbers of input points ($251\times251$ mesh, $\zeta=50$)}
	\label{table:franke}
	\begin{tabular}{L{1.2cm} C{2.1cm} C{3.45cm} C{2.1cm} C{3.45cm}}
		\toprule
		\multicolumn{1}{c}{} & $E\left(W^{\text{MMS}}_{\text{Franke}}\right)$ & $I\left(W^{\text{MMS}}_{\text{Franke}}, W_{\text{Franke}}\right)$ & $E\left(W^{\text{MMS}}_{\text{Franke}}\right)$ & $I\left(W^{\text{MMS}}_{\text{Franke}}, W_{\text{Franke}}\right)$ \\ \midrule
		BCs & Correct & Correct & Free & Free\\
		\midrule
		\multicolumn{5}{l}{\textbf{Regular Distribution}}\\
		$3 \times 3$   & 57.1820  & $9.48 \cdot 10^{-3}$ & 0.6300   & $4.11 \cdot 10^{-2}$ \\
		$7 \times 7$   & 172.5891 & $1.71 \cdot 10^{-4}$ & 159.3226 & $2.68 \cdot 10^{-4}$ \\
		$11 \times 11$ & 205.5717 & $6.0 \cdot 10^{-6}$  & 201.0426 & $8.0 \cdot 10^{-6}$  \\
		$25 \times 25$ & 213.9168 & $< 10^{-6}$          & 212.3111 & $< 10^{-6}$          \\ \midrule
		\multicolumn{5}{l}{\textbf{Uniform Distribution}}\\
		10   & 84.0491  & $3.55 \cdot 10^{-3}$ & 24.7928  & $2.79 \cdot 10^{-2}$ \\
		25   & 135.2511 & $1.65 \cdot 10^{-3}$ & 54.6206  & $1.01 \cdot 10^{-2}$ \\
		50   & 144.6379 & $9.45 \cdot 10^{-4}$ & 121.2439 & $1.93 \cdot 10^{-3}$ \\
		100  & 196.1192 & $7.3 \cdot 10^{-5}$  & 181.9452 & $2.11 \cdot 10^{-4}$ \\ \midrule
		\textbf{Exact}  & \textbf{214.1254} & \textbf{0} & \textbf{214.1254} & \textbf{0} \\ \bottomrule
	\end{tabular}
\end{table}

It is worth noting that regular grids generally outperform random sets of comparable size (e.g., 49 grid points vs. 50 random ones) under correct boundary conditions: the energies are 172.59 and  144.64, respectively, since regularly spaced points sample the domain more evenly. Nevertheless, random distributions also produce physically meaningful surfaces and demonstrate the robustness of the method with respect to the arrangement of input data.

The two metrics together reveal an important property of the MMS reconstruction. The relative deficit of the surface energy with respect to $E\left(W_{\text{Franke}}\right)=214.1254$ is consistently much larger than the relative geometric deviation of the surface from the target, measured by the relative error of the displacement $\sqrt{\frac{I\left(W^{\text{MMS}}_{\text{Franke}}, W_{\text{Franke}}\right)}{I(W_{\text{Franke}})}}$. For example, at the $11\times 11$ regular grid with correct boundary conditions, $E\left(W^{\text{MMS}}_{\text{Franke}}\right)$ lies about 4.0\% below $E(W_{\text{Franke}})$, whereas the relative error of the displacement is only about 0.49\%. The same pattern persists across the table: even when the energy is noticeably below the original (which corresponds to a smoother, more biharmonic surface), the surface itself remains a high-fidelity geometric approximation of the target. The gain in smoothness offered by MMS is therefore not paid for by a comparable geometric deviation. In the context of our broader methodology, this confirms that the MMS successfully bridges the disciplinary divide: it delivers the high-order aesthetic fairness required by CG communities while strictly maintaining the mathematical and physical accuracy expected in CM.

\begin{figure}[ht!]
	\centering
	\begin{subfigure}[b]{0.48\textwidth}
		\centering
		\includegraphics[width=\textwidth]{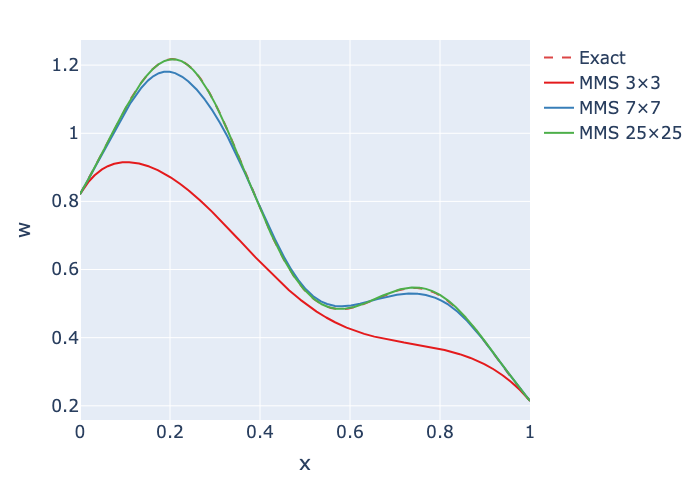}
		\caption{Surface elevation}
	\end{subfigure}
	\hfill
	\begin{subfigure}[b]{0.48\textwidth}
		\centering
		\includegraphics[width=\textwidth]{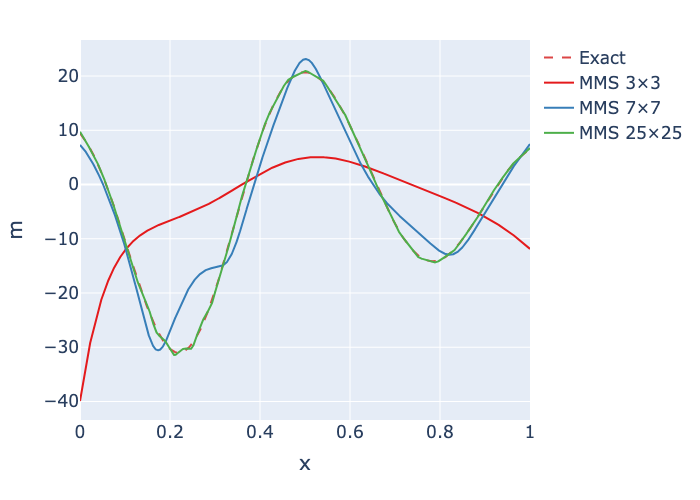}
		\caption{Curvatures}
	\end{subfigure}
	\caption{Franke's function results along the line $y\approx 0.22$ ($251 \times 251$, correct boundary conditions, $\zeta=50$)}
	\label{fig:franke_results}
\end{figure}

Figure \ref{fig:franke_results} demonstrates the displacement and moment profiles along the line $y\approx0.22$ (passing through the region of the dominant peak) under correct boundary conditions for three regular grids of input points. With $3\times3$ points, the MMS curve captures only the coarse shape and strongly smooths out both the peak and the moment oscillations. Already at $7\times7$ points, the reconstruction matches the exact displacement almost everywhere, with only minor deviations in the moments near the peak. At $25\times25$ points, both the displacement and moments visually coincide with the exact solution.

\section{Conclusion}
\label{sec:conclusion}

While computational mechanics and computer graphics share roots in the physics of thin plates, their numerical objectives often diverge. CM often prioritizes nodal accuracy, whereas CG demands global smoothness and curvature continuity. This paper bridges these fields using the Method of Matched Sections. Unlike standard conforming Finite Element Method, which may violate higher-order continuity at element boundaries, MMS results in a visually plausible representation, suitable for graphics applications.

We demonstrated that MMS:
\begin{enumerate}
	\item Successfully incorporates corner constraints and point data into the variational formulation. As demonstrated with the example of a point-supported surface, this novel approach to handling concentrated constraints led to a significant improvement in accuracy. This capability is a key step in reconciling the discrete nature of mechanical constraints with the continuity requirements of computer graphics.
	
	\item Accurately reconstructs surfaces from diverse boundary data (derivatives of 1st or 2nd order).  Numerical examples show rapid and precise convergence to the exact surface, confirming that our mathematical formulation is ideally suited for surface blending tasks where seamless continuity between different patches is paramount.
	
	\item Generates surfaces that are inherently biharmonic and energetically optimal, surpassing the fairness of the generating functions. When boundary conditions are derived from an arbitrary, non-biharmonic function, the surface generated by our method is inherently biharmonic. This resulting surface is not only smooth but also energetically optimal, possessing a lower thin-plate energy value than the original function from which the boundaries were taken. If an internal measurement point is added as a constraint, the reconstructed surface becomes geometrically closer to the original generating surface. Even in this constrained state, the surface's energy characteristics remain superior to those of the original, demonstrating an intrinsic tendency towards the smoothest possible shape that honors the given data.
	
	\item Through a local regularization kernel, the method smooths curvature singularities induced by point constraints, ensuring visual quality for design applications. This approach successfully restores the continuity of curvatures while only marginally increasing the total surface energy, offering a practical way to regularize the solution for visual fairness.
\end{enumerate}

\section*{Credit authorship contribution statement}
\textbf{Igor Orynyak}: Conceptualization, Methodology, Writing --- original draft.
\textbf{Kirill Danylenko}: Software, Validation, Visualization, Writing --- review \& editing.
\textbf{Danylo Tavrov}: Validation, Writing --- review \& editing.

\section*{Declaration of competing interest}
The authors declare that they have no known competing relationships that could have appeared to influence the work reported in this paper.

\section*{Acknowledgments}
The second author acknowledge the partial financial support provided by Simulmedia Inc LLC. The sponsor had no role in the study design, data analysis, or the decision to publish the results.

\appendix
\section{Analytical Solutions for the 1D Directional Components}
\label{sec:appendix_derivations}

By integrating the coupled ordinary differential equations presented in Section \ref{sec:mms_theory}, we obtain the dependencies of the parameters along the central lines. For the third derivatives along $x$:
\begin{equation}
	Q^x(x) = Q^x_0 + A_1 x\;,
	\label{eq:force_x}
\end{equation}
where the subscript 0 indicates the initial section, so $Q^x_0 = Q^x(0)$. For the third derivative $Q^y(y)$ along $y$:
\begin{equation}
	Q^y(y) = Q^y_0 + (P - A_1) y\;,
	\label{eq:force_y}
\end{equation}
where $P$ is the distributed data term (or external load). In geometrical applications, it can be used to account for given geometrical constraints. For the twisting second derivatives:
\begin{align}
	M_{\tau}^y(y) &= M_{\tau,0}^y + A_2 y\;, \label{eq:twist_moment_y} \\
	M_{\tau}^x(x) &= M_{\tau,0}^x + A_3 x\;. \label{eq:twist_moment_x}
\end{align}
For the bending second derivatives:
\begin{align}
	M_{n}^x(x) &= M_{n,0}^x + Q_{0}^x x + \frac{A_1 x^2}{2} - A_2 x\;, \label{eq:bend_moment_x} \\
	M_{n}^y(y) &= M_{n,0}^y + Q_{0}^y y + \frac{(P-A_1)y^2}{2} - A_3 y\;. \label{eq:bend_moment_y}
\end{align}
The integrally averaged curvature parameters are:
\begin{align}
	M_{n,\text{avg}}^x &= \frac{1}{a}\int_0^a M_{n}^x(x)\, dx = M_{n,0}^x + \frac{Q_{0}^xa}{2} + \frac{A_1 a^2}{6} - \frac{A_2 a}{2}\;, \label{eq:avg_bend_moment_x} \\
	M_{n,\text{avg}}^y &= \frac{1}{b}\int_0^b M_{n}^y(y)\, dy = M_{n,0}^y + \frac{Q_{0}^y b}{2} + \frac{(P-A_1)b^2}{6} - \frac{A_3 b}{2}\;. \label{eq:avg_bend_moment_y}
\end{align}

Let $D$ denote the flexural rigidity of the plate. From the geometric equations for the twist angles:
\begin{align}
	\theta_{\tau}^x(x) &= \theta_{\tau,0}^x + \frac{M_{\tau,0}^x}{D(1-\nu)}x + \frac{A_3}{D(1-\nu)}\frac{x^2}{2}\;, \label{eq:twist_angle_x} \\
	\theta_{\tau}^y(y) &= \theta_{\tau,0}^y + \frac{M_{\tau,0}^y}{D(1-\nu)}y + \frac{A_2}{D(1-\nu)}\frac{y^2}{2}\;. \label{eq:twist_angle_y}
\end{align}
The solutions for slope angles are given by:
\begin{align}
	\theta_{n}^x(x) &= \theta_{n,0}^x + \frac{1}{D(1-\nu^2)} \left( M_{n,0}^x x + \frac{Q_{0}^x x^2}{2} + \frac{A_1 x^3}{6} - \frac{A_2 x^2}{2} \right) \nonumber \\
	&\quad - \frac{\nu}{D(1-\nu^2)} M_{n,\text{avg}}^y \cdot x\;, \label{eq:bend_angle_x} \\
	\theta_{n}^y(y) &= \theta_{n,0}^y + \frac{1}{D(1-\nu^2)} \left( M_{n,0}^y y + \frac{Q_{0}^y y^2}{2} + \frac{(P-A_1)y^3}{6} - \frac{A_3 y^2}{2} \right) \nonumber \\
	&\quad - \frac{\nu}{D(1-\nu^2)} M_{n,\text{avg}}^x \cdot y\;. \label{eq:bend_angle_y}
\end{align}
Finally, integration yields the surface elevation functions:
\begin{align}
	w^x(x) &= w_{0}^x + \theta_{n,0}^x x + \frac{1}{D(1-\nu^2)} \left( \frac{M_{n,0}^x x^2}{2} + \frac{Q_{0}^x x^3}{6} + \frac{A_1 x^4}{24} - \frac{A_2 x^3}{6} \right) \nonumber \\
	&\quad - \frac{\nu}{D(1-\nu^2)} M_{n,\text{avg}}^y \cdot \frac{x^2}{2}\;, \label{eq:displacement_x} \\
	w^y(y) &= w_{0}^y + \theta_{n,0}^y y + \frac{1}{D(1-\nu^2)} \left( \frac{M_{n,0}^y y^2}{2} + \frac{Q_{0}^y y^3}{6} + \frac{(P-A_1)y^4}{24} - \frac{A_3 y^3}{6} \right) \nonumber \\
	&\quad - \frac{\nu}{D(1-\nu^2)} M_{n,\text{avg}}^x \cdot \frac{y^2}{2}\;. \label{eq:displacement_y}
\end{align}

\bibliography{references}
	
\end{document}